\documentclass[aps,prd,10pt,twocolumn,showpacs,preprintnumbers,amsmath,amssymb,nofootinbib]{revtex4}

\usepackage{natbib} 

\usepackage{color}

\usepackage{amsmath}
\usepackage{epsfig}
\usepackage{array}
\usepackage[hyperfootnotes=true]{hyperref} 
\usepackage{afterpage}
\usepackage{color}

\begin{document}

\title{Backward-Propagating MeV Electrons in Ultra-Intense Laser Interactions: \\ Standing Wave Acceleration and Coupling to the Reflected Laser Pulse}

\author{  Chris Orban$^{1,2,*}$, John T. Morrison$^{3,5}$, Enam A. Chowdhury$^{1,4}$,  John A. Nees$^{2,5}$, Kyle Frische$^{2}$, Scott Feister$^{1,2}$ and W. M. Roquemore$^{6}$}

\affiliation{
\vspace{0.2cm}
(1) Department of Physics, The Ohio State University, Columbus, OH \\
(2) Innovative Scientific Solutions, Inc., Dayton, OH \\
(3) Fellow, National Research Council \\
(4) Intense Energy Solutions, Inc., Dayton, OH \\
(5) Center for Ultra-Fast Optical Science, University of Michigan, Ann Arbor, MI \\
(6) Air Force Research Laboratory, Dayton, OH 
}

\email{orban@physics.osu.edu}

\date{\today}

\begin{abstract}
Laser-accelerated electron beams have been created at a kHz repetition rate from the {\it reflection} of intense ($\sim10^{18}$~W/cm$^2$), $\sim$40~fs laser pulses focused on a continuous water-jet in an experiment at the Air Force Research Laboratory. This paper investigates Particle-in-Cell (PIC) simulations of the laser-target interaction to identify the physical mechanisms of electron acceleration in this experiment. We find that the standing-wave pattern created by the overlap of the incident and reflected laser is particularly important because this standing wave can ``inject'' electrons into the reflected laser pulse where the electrons are further accelerated. We identify two regimes of standing wave acceleration: a highly relativistic case ($a_0~\geq~1$), and a moderately relativistic case ($a_0~\sim~0.5$) which operates over a larger fraction of the laser period. In previous studies, other groups have investigated the highly relativistic case for its usefulness in launching electrons in the forward direction. We extend this by investigating electron acceleration in the {\it specular  (back reflection) direction} and over a wide range of intensities ($10^{17}-10^{19}$ W cm$^{-2}$). 
\end{abstract}

\maketitle

\section{Introduction}
  \label{sec:intro}

Laser-accelerated electron beams from ultra-intense laser-matter interactions  with solid targets have been observed and studied in diverse contexts. This has been a topic of great interest in part because of the potential for these electrons to deliver energy to the compressed core of an inertial confinement fusion target \citep{Tabak_etal1994,AkliOrban_etal2012}. Other important applications for laser-accelerated electron beams include radiotherapy with electron energies well above what can be achieved with conventional linear accelerators \cite{Glinec_etal2006,Fuchs_etal2009}. The creation of highly pulsed x-ray and UV radiation through Compton scattering of another pulse of light with the electron beam presents another unique opportunity. While there has been much success in Compton scattering off electron beams accelerated from underdense targets, e.g., from laser wakefield acceleration \cite{Powers_etal2014}, \citet{Naumova_etal2004} highlight the possibilities that sub-fs bunches of electrons from laser-solid interactions offer for creating Compton light sources with extremely short timescale pulsing. Using ultra-intense, ultra-short lasers to create these electron beams can significantly shorten the temporal duration of the resulting electron pulses, even to timescales shorter than what can be achieved with photocathode technology \cite{Bentson_etal2004}. Thus laser-accelerated electron bunches may provide some advantage to efforts to create free electron lasers \cite{Hooker2013}. A number of groups are investigating laser-accelerated electron sources with these goals in mind. 

The next section describes measurements of laser-accelerated electrons and dosiometric measurements of bremsstrahlung radiation from a laser experiment conducted at the Air Force Research Laboratory (AFRL) at Wright-Patterson Air Force Base in Dayton, Ohio. As described in the next section and in a follow up paper by Morrison et al. \cite{Morrison_etal2015} ultra-intense laser interactions at normal incidence are found to produce significant radiation in the specular (back reflection) direction in spite of the tendency for $\vec{J} \times \vec{B}$ forces to accelerate electrons in the forward direction. As discussed in Morrison et al. \cite{Morrison_etal2015}, the total charge in relativistic electrons is of order 0.3~nC, which is substantially more charge than comparable laser wakefield experiments.

This paper presents the first Particle-in-Cell (PIC) simulations of the water-jet target experiment just mentioned. These simulations were performed using the LSP code \citep{Welch_etal2004} and electron trajectories were followed in detail in order to understand the precise mechanisms involved in the creation of the electron beam. While a substantial literature of simulation/modeling papers exists that seeks to understand forward-going electron acceleration from laser interaction with solid targets \citep[e.g.][]{Pegoraro_etal1997,Kemp_etal2009,May_etal2011,Ovchinnikov_etal2013}, and a number of other papers investigate electron acceleration from obliquely incident laser light interacting with solids \citep{Brunel1988,Ruhl_etal1999,Cai_etal2003,Cai_etal2004,Chen_etal2006,Habara_etal2006,Li_etal2006,Brandl_etal2009,Wang_etal2010,Tian_etal2012,SanyasiRao_etal2012,Perez_etal2013}, significantly less attention has been devoted to specularly-directed electron acceleration mechanisms near normal incidence. Our goal is to understand the mechanisms involved with specularly-accelerated electrons in a qualitative way and to help fill this gap in the literature. These insights into the mechanism will ultimately create a foundation for further optimization of the electron beam parameters (energies, total charge, emittance) in this and related experiments. 

\S~\ref{sec:exp} describes the experiment at AFRL and discusses two salient results that motivate this paper. Sec.~~\ref{sec:sims} describes the PIC simulations using the LSP code to understand the ultra-intense pulse interaction with the target. In Sec.~\ref{sec:results} the results from these simulations and particle tracking are considered. Sec.~\ref{sec:mech} describes the highly relativistic and moderately relativistic regimes of standing wave acceleration. Sec.~\ref{sec:conclusions} summarizes our results and main conclusions. Finally, Appendix~\ref{ap:thresh} provides some approximate analytic insights into standing wave acceleration.

\section{Experimental Setup}
\label{sec:exp}

The motivation of this work comes from an experiment performed at the Air Force Research Laboratory at Wright-Patterson Air Force Base in Dayton, Ohio. The experimental setup and and some salient observations are briefly described here. Further experimental details and results are described in Morrison et al. \cite{Morrison_etal2015}.

The experiment was carried out with a modified Red Dragon short pulse laser system \cite{kmlabs}.  Laser pulses of 35~fs FWHM duration and 3~mJ of energy are focused by a metallic off-axis parabolic (OAP) mirror in f/1.3 configuration at normal incidence onto a 30 $\mu$m diameter flowing water column. The experiment is conducted in a vacuum chamber held at 20~Torr background pressure in order to avoid freezing the water jet flow. 

The peak laser intensity in this experiment is near $10^{18}$~W/cm$^2$. Since the experiment is conducted at 20~Torr and not high vacuum, the effect of laser focus distortion (self-focusing) due to the intensity dependence of the index of refraction must be assessed. Analytically, this effect can be quantified through computing the ``B-integral'' for the laser pulse \cite{Paschotta2008}. Using the experimental parameters, the B-integral is estimated to be less than 0.2 for the laser propagating between the OAP up to 25~$\mu$m away from the water jet target, which is approximately where the ultra-intense pulse may encounter significant pre-plasma. As a rule of thumb, self-focusing only becomes important for B-integral values of 3-5 or more \cite{Paschotta2008}. Thus the effect is likely to be small.

This conclusion is also supported by empirical evidence from a frequency-doubled 400~nm probe pulse of the laser plasma interaction. This diagnostic provides shadowgraphy and interferometry of the interaction region on a femtosecond-to-ns time delay \cite{Feister_etal2014}. Experiments in air with full laser pulse energy and without a water jet target show that plasma channel formation does not occur below a 150~Torr threshold according to both shadowgraphy and interferometry. This result provides empirical evidence that the effect of self-focusing is small at the 20~Torr operating pressure of the experiment.

\begin{figure}
\includegraphics[angle=0,width=3.2in]{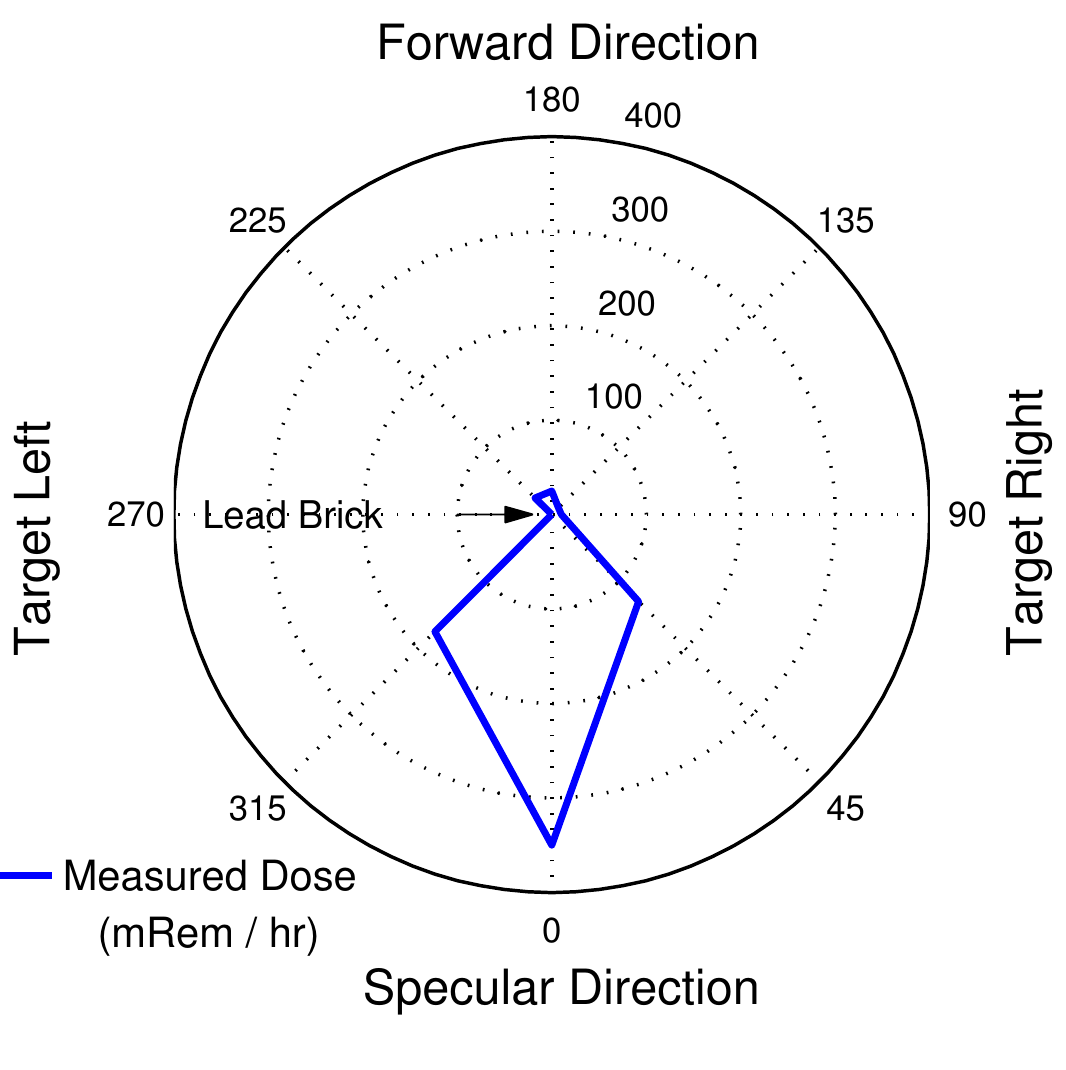}
\vspace{-0.5cm}
\caption{Radiation produced by ultra-intense laser-matter interactions as measured by a dosimeter placed at various locations outside the target chamber. The experimental setup, in which an ultra-intense laser pulse is normally incident on a water jet target, is described in Fig.~\ref{fig:lanex}. The dosimeter measurements, which are sensitive to x-ray energies $>$25~keV, show that most of the radiation is produced in the ``backwards'' direction, opposite to the laser propagation direction. As indicated in the diagram, a lead brick placed \emph{inside} the target chamber significantly attenuates the measured signal in one particular direction \emph{outside} the target chamber. None of the other measurements are attenuated in this way.}\label{fig:dose}
\end{figure}

X-ray emission was monitored outside the chamber by a radiation survey meter (Fluke Biomedical, Model 451P, Ion Chamber Survey Meter) sensitive to x-rays above 25~keV. It was discovered that the radiation dose is unusually high outside the chamber behind the OAP compared to the forward propagation direction (directly opposite side), and all other angular positions, as shown in a polar plot in Fig.~\ref{fig:dose}. This was completely unexpected from prior experiences with ultra-intense laser plasma interactions, where the $\vec{J} \times \vec{B}$ force dominates to push electrons forward, resulting in a bremsstrahlung radiation peaked in the forward direction. We hypothesized that the back-directed radiation dose could be explained if a significant number of energetic electrons generated during the laser plasma interaction propagated backward, generating bremsstrahlung radiation in the aluminum OAP, its mount, and the chamber wall.

\begin{figure}
\includegraphics[angle=0,width=3.4in]{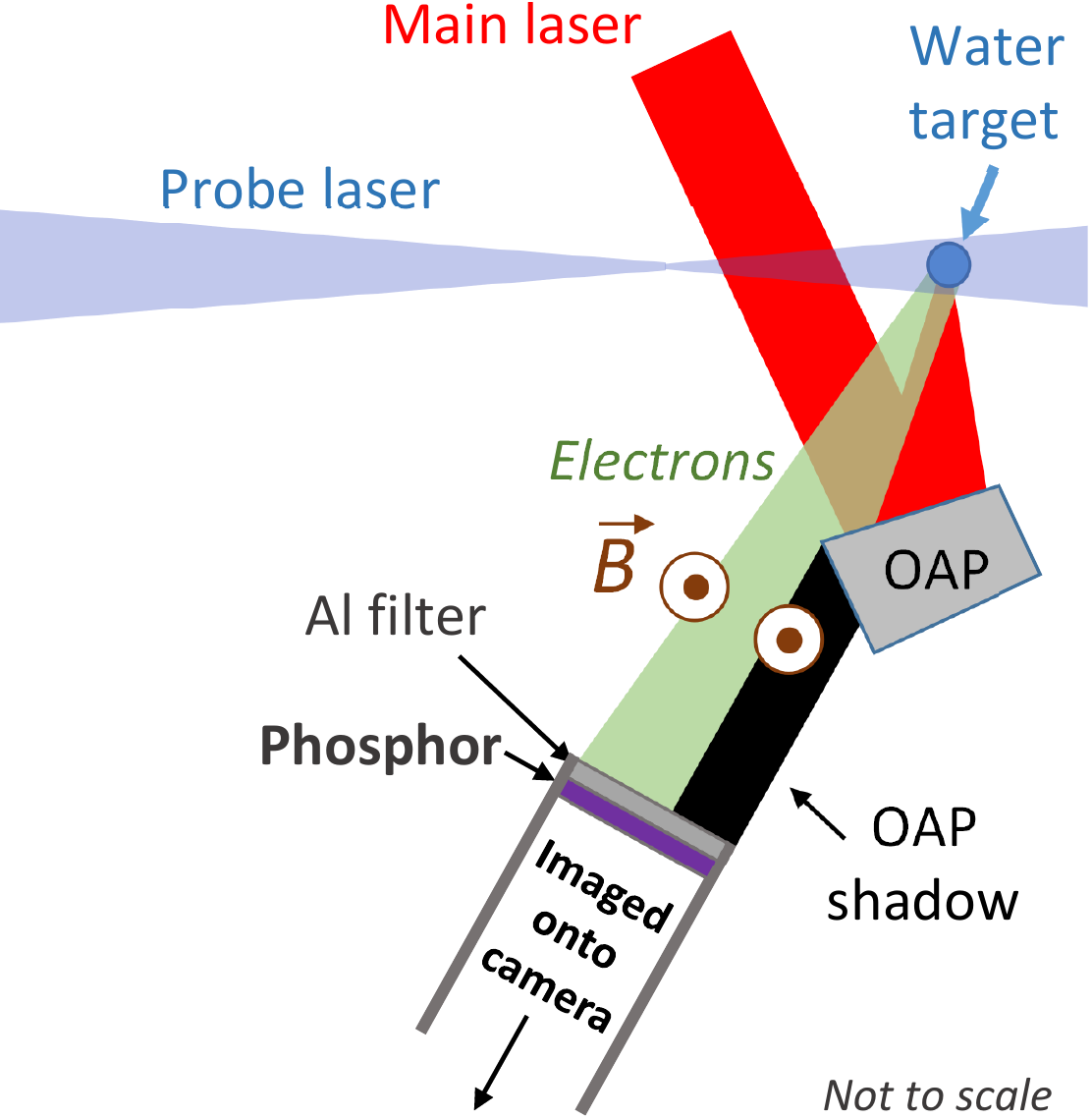}
\vspace{-0.3cm}
\caption{The experimental setup is shown in which the main laser pulse (800~nm, 3~mJ) is focused by an off-axis parabola (OAP) onto a water jet target. A frequency-doubled probe pulse at 400 nm is used to obtain shadowgraphy and interferometry of the target region \cite{Feister_etal2014}. Also shown is a diagnostic where energetic electrons ejected from the water jet are incident on a fluorescent ``Lanex'' screen \cite{Wagner_etal1997}. A camera in a light-tight housing images the optical light produced by the screen (Fig.~\ref{fig:results}). Measurements were made with and without a rare earth magnet ($\sim$0.16~T surface field).}\label{fig:lanex}
\end{figure}

\begin{figure*}
\includegraphics[angle=0,width=6in]{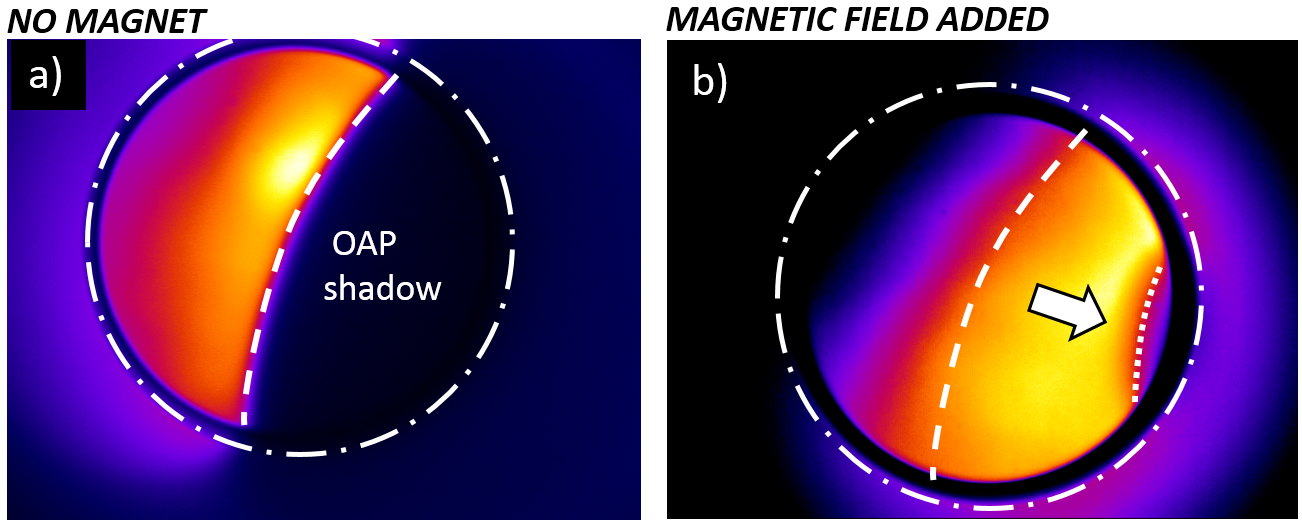}
\caption{False-color visible-light images of fluorescent emission from a Lanex screen due to energetic electrons arriving from the target region. The edge of the Lanex screen is shown by a dash-dotted white line. Light recorded outside of this is due to scatter from the beam tube. Panel a) presents results without a rare earth magnet, showing a beam-like feature on the screen and a distinct shadow created by the off-axis parabola (OAP) which was highlighted earlier in Fig.~\ref{fig:lanex}. Panel b) presents results when a rare earth magnet is added as in Fig.~\ref{fig:lanex}. In this panel the OAP shadow moves from its original position (dashed white line) to significantly further to the right (dotted white line), validating the hypothesis that the fluorescence is due to energetic electrons.}\label{fig:results}
\end{figure*}

To test this hypothesis, a 1~inch diameter Lanex screen \cite{Wagner_etal1997} was positioned behind the OAP to directly observe the backward-propagating electrons. This setup is shown in Fig.~\ref{fig:lanex}. A 25~$\mu$m thick aluminum foil placed in front of the Lanex renders the setup light tight and also acts as low energy electron filter (blocks most $\lesssim$~50~keV electrons). When energetic electrons hit the Lanex screen, light is emitted in the visible spectrum. This light is imaged onto a 12~bit CCD camera outside the vacuum chamber. Fig.~\ref{fig:results} (a) shows phosphor images of backward propagating electrons partially obscured by the OAP hitting the Lanex detector. Fig.~\ref{fig:results} (b) shows the same view when a single rare earth magnet (0.16~T surface field) was placed just above the OAP.  The direction of deflection in the presence of this magnetic field confirms that the fluorescence is due to negatively charged electrons. Electrons are deflected and dispersed on the Lanex screen, with the OAP shadow moving $\sim$12~mm. Note that the distance between the OAP and the water jet target is only $\sim$27~mm, making it difficult to place a pair of magnets with a yoke in this space. 

Since the backward-directed radiation source is deflected in a manner consistent with electrons and transmits through an aluminum filter blocking $\gtrsim 50$~keV electrons, we conclude the presence of backward-propagating energetic electrons originating from the laser plasma interaction. Further investigation of these laser-accelerated electrons is presented in \cite{Morrison_etal2015}. { Evidence is presented there that indicates of order 0.3~nC of relativistic electrons are being accelerated per shot.}

\section{Particle-In-Cell Simulations}
\label{sec:sims}

\subsection{Simulation Setup}

\begin{figure*}
\includegraphics[angle=0,width=6.5in]{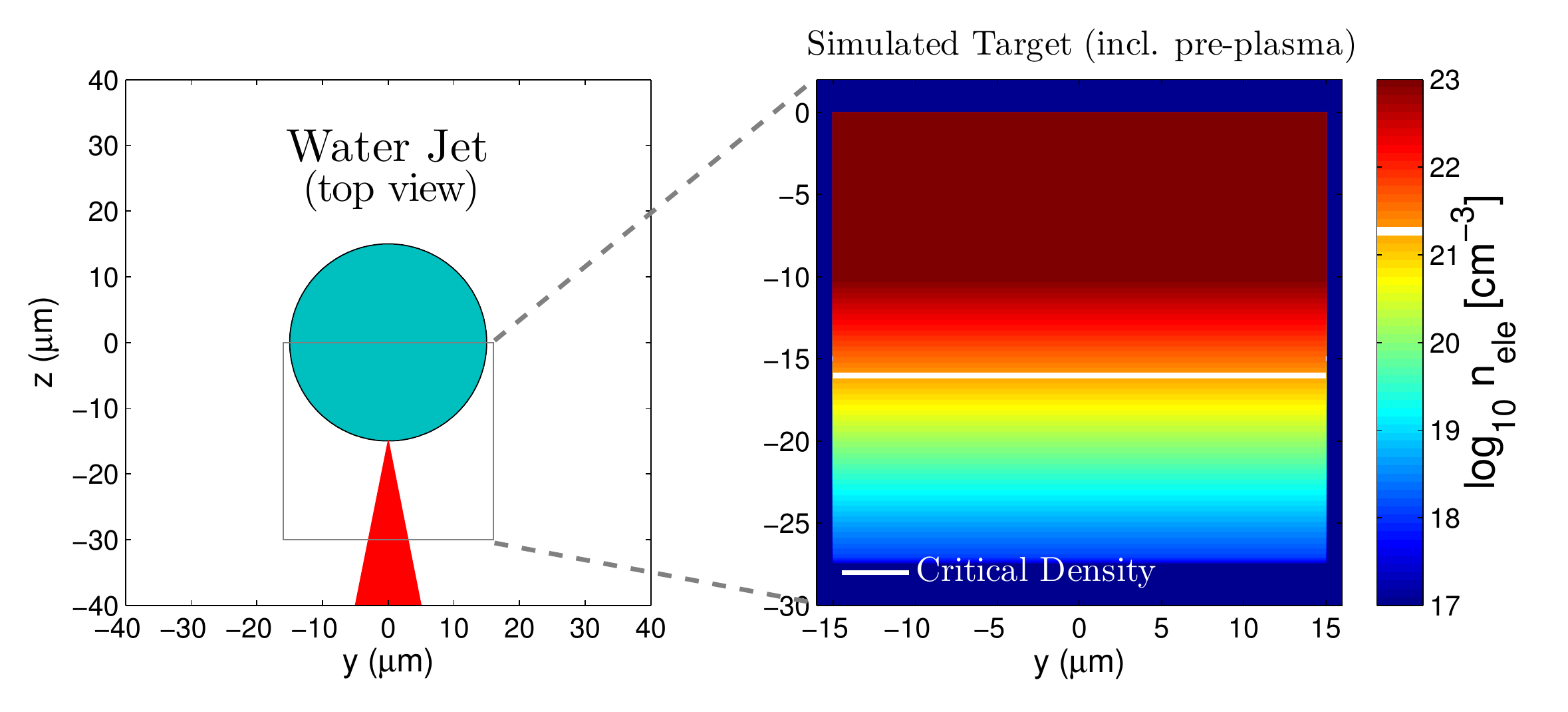}
\vspace{-0.5cm}
\caption{\emph{Left panel}: A top-down illustration of the 30 $\mu$m diameter water-jet target used in the experiment. The laser polarization is in the $y$-direction, and the pulse converges on the surface of the water with an $f$-number of 2.5. \emph{Right panel}: The electron number density for the simulated target including a 1.5~$\mu$m scale length pre-plasma. As illustrated by the dashed gray lines between the left and right panels, the simulated target is smaller than the real target and, for simplicity, the target is flat. A thick white line indicates the location of the (non-relativistic) critical density for electrons in 800 nm light ($n_{\rm crit} = 1.72 \cdot 10^{21}$~cm$^{-3}$). Simulations were performed with a 2D(3$v$) cartesian geometry using the Particle-in-Cell code LSP.}
\label{fig:lsptarget}
\end{figure*}

We performed PIC simulations using the code LSP \cite{Welch_etal2004} in a 2D(3$v$) Cartesian geometry. Since the laser-target interaction is simulated using only two spatial dimensions, symmetry must be assumed along some physical axis. We take advantage of the translational symmetry along the length of the water jet and make the natural choice to set the symmetry direction parallel to the direction of the jet as illustrated by the left panel of Fig.~\ref{fig:lsptarget}. The simulated laser pulse strikes the water target with a tangential polarization vector as it does in the experiment. 

For simplicity and in an effort to minimize the expense of the simulations we do not simulate the entire~30 $\mu$m diameter of the water jet. Instead, as illustrated by the gray box in the left panel of Fig.~\ref{fig:lsptarget}, only a section of the water jet is simulated. This section is further simplified to a flat slab geometry (instead of including the natural curvature of the water jet) as shown in the right panel of Fig.~\ref{fig:lsptarget}. Because of the smallness of the laser spot size ($1.5~\mu$m FWHM) relative to the $30~\mu$m diameter of the water jet, our qualitative conclusions are unchanged whether a realistic target curvature or a slab geometry is used.

An important aspect of the experiment is the presence of a pre-plasma that extends many microns away from the edge of the water jet. As discussed in the next section, the presence or absence of a pre-pulse determines whether or not an electron beam will be created in the main pulse interaction. As shown in Fig.~\ref{fig:lanex}, a 400~nm probe pulse provides shadowgraphy and interferometry of the laser-interaction region. { By operating at 400~nm the probe transmits through up to four times the critical density of the 800~nm main pulse.} Feister et al. \cite{Feister_etal2014} describe this setup and the data presented there indicates the presence of a pre-plasma extending $\sim$10-20~$\mu$m from the target. Since the purpose of the present investigation is to understand the mechanism of electron acceleration, for simplicity we assume an exponential scale length pre-plasma density profile instead of using a pre-plasma profile directly inferred from interferometry. To be consistent with the overall extent of the observed pre-plasma, this exponential scale length must be $\ll$10-20~$\mu$m. The exponential scale length is set to be 1.5~$\mu$m throughout this paper. Our qualitative results are unchanged for scale lengths as small as $\sim$0.75~$\mu$m and as large as $\sim$4~$\mu$m. In future work the shadowgraphy and the interferometry data will be used to make the simulated pre-plasma density profiles significantly more realistic.

Simulations were performed on the Spirit supercomputer using $\lambda$/32~$\times~\lambda/32$~= 0.025~$\mu$m~$\times$~0.025~$\mu$m resolution and $\Delta t=0.05$~fs timesteps, which is over 50 timesteps per laser cycle. Each cell with non-zero density is assigned 49 electron macroparticles, 49 singly ionized oxygen macroparticles and 49 proton macroparticles\footnote{n.b. $49 = 7^2$. It is most convenient in LSP to specify an integer squared number of particles per cell.}. The charge for each macroparticle was set to keep each cell initially charge neutral while assuming a mixture of two parts ionized hydrogen and one part singly-ionized oxygen. During the course of the simulation the laser pulse can further ionize the oxygen ions through field ionization according to the Ammisov-Delone-Krainov rate \citep{ADK}. Although we do not expect the real water jet to become fully singly-ionized by the pre-pulse, the rapid timescale of ionization by the intense laser fields in the simulation implies that the results should be insensitive to the precise state of the target and pre-plasma at the beginning of the simulation.

A Monte-Carlo scattering algorithm was applied each timestep to the electrons in the simulation \citep{Kemp_etal2004}. The scattering rate came from the classical Spitzer formula \cite{Atzeni2004} except at very low temperatures where the scattering rate was bounded by the finite timestep of the simulations ($\Delta t^{-1} = 2 \cdot 10^{16}$~Hz).

The simulations were run over 500~fs in order to adequately model the laser propagation, target interaction and the propagation of the electron beam. Electron trajectories were tracked in these simulations for later analysis (Sec.~\ref{sec:results}).

\subsection{Intensities, Focus and Spot Size}

Since the goal of the present study is to qualitatively understand the mechanisms of electron acceleration we investigate in detail three different peak intensities ($5 \cdot 10^{17}$, $10^{18}$, and $5 \cdot 10^{18}$~W~cm$^{-2}$) with the same gaussian spot size (1.5 $\mu$m FWHM, or $f=2.5$) and the same duration (sine squared envelope with 30 fs FWHM). These intensities are summarized in Table~\ref{tab:intensities}. As mentioned in Sec.~\ref{sec:intro}, the peak intensity in the experiment is estimated to be $3 \cdot 10^{18}$ W cm$^{-2}$ \cite{Morrison_etal2015}. Thus the ``Moderately Relativistic Case'' is closest to the experimental conditions and the ``Relativistic Case'' is significantly more intense than the experiment. The ``Mildly Relativistic Case'' provides an interesting comparison where, as will be discussed in the next section, electron acceleration in the specular direction is less efficient than at higher intensities.

\begin{table}
\begin{tabular}{| l | c | c | c | }
\hline
& $I_{\rm peak}$ & $a_{\rm peak}$ & $a_{\rm SW}$  \\ \hline
Mildly Relativistic Case & $ 5 \cdot 10^{17}$ W cm$^{-2}$ & 0.48 & 0.39 \\ \hline
Moderately Relativistic Case & $10^{18}$ W cm$^{-2}$ & 0.68 & 0.55 \\ \hline
Relativistic Case & $5 \cdot 10^{18}$ W cm$^{-2}$ & 1.5 & 1.24 \\ \hline
\end{tabular}
\caption{Simulated peak intensities and $a$-values for freely propagating beam ($I_{\rm preak}$, $a_{\rm peak}$) and the peak $a$-value for the standing wave pattern ($a_{\rm SW}$)} \label{tab:intensities}
\end{table}

The peak intensities, $I_{\rm peak}$, listed in Table~\ref{tab:intensities} should be understood as the peak intensity that would be achieved at peak focus in vacuum (i.e. \emph{without} a target). Similar to the experiment, the critical density in our simulations (white line in Fig.~\ref{fig:lsptarget}) is placed about 3 rayleigh lengths in front of where the peak focus would be (i.e. at the origin, $z = y = 0~\mu$m in Fig.~\ref{fig:lsptarget}). This choice would imply that the intensities achieved in the simulation remain significantly lower than $I_{\rm peak}$, however the standing wave pattern created by the incident and reflected pulses creates constructive interference that increases the maximum intensity. Ultimately the typical intensity of the standing wave pattern in the simulation is similar to $I_{\rm peak}$. We report the $a$-value for this typical intensity of the standing wave ($= a_{\rm SW}$) in the far-right column of Table~\ref{tab:intensities}. 

Finally, note that LSP assumes a perfectly-gaussian laser pulse with a diffraction limited spot size (1.5 $\mu$m FWHM). The spot size in the experiment will typically be somewhat larger than this ideal ($\approx 2.2~\mu$m FWHM from Morrison et al. \cite{Morrison_etal2015}). For the PIC simulations we adopt the ``ideal'' value of 1.5~$\mu$m FWHM, but our essential conclusions do not sensitively depend on the choice of spot size.

\begin{figure*}
\centerline{\includegraphics[angle=0,width=4in]{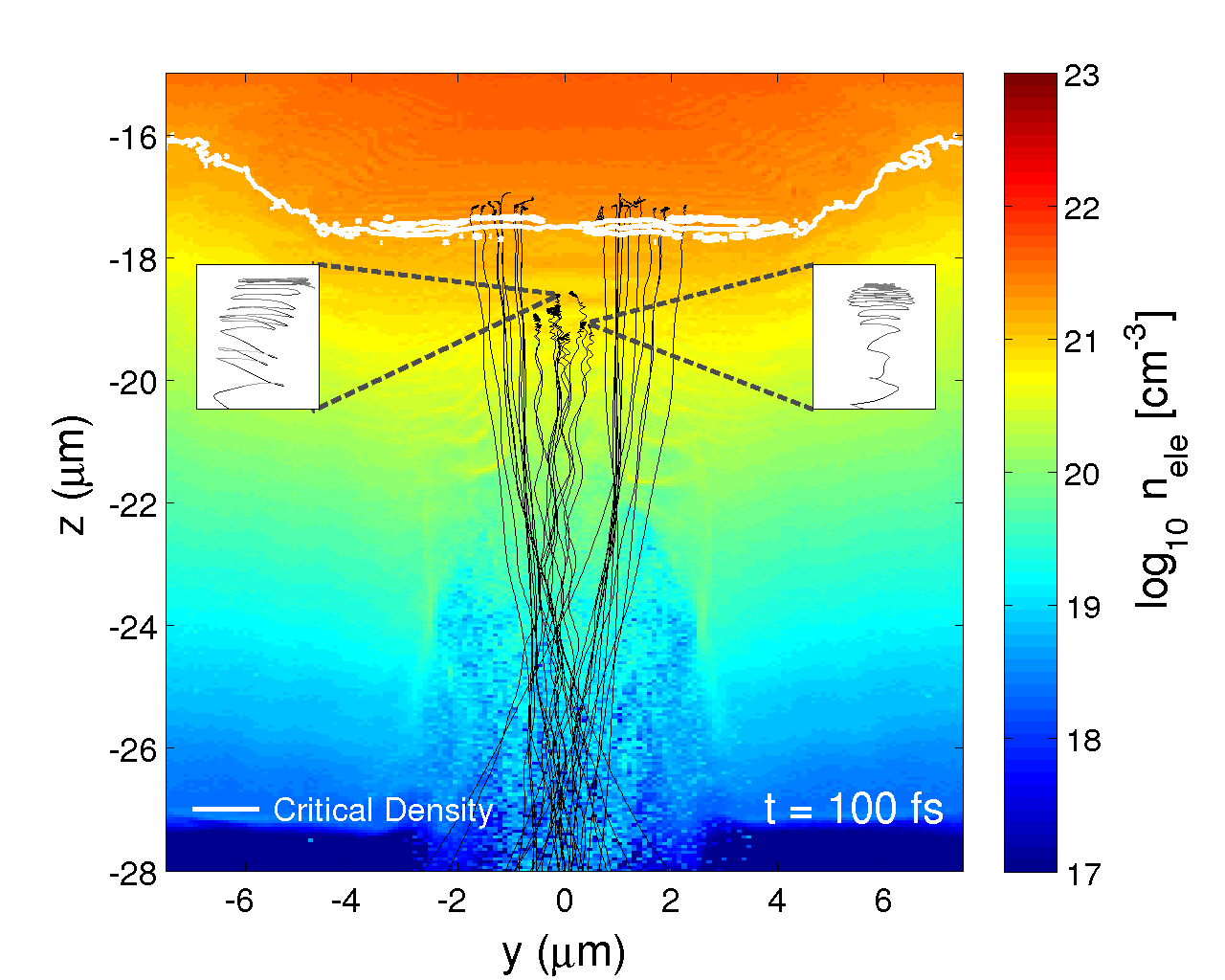} \, \, \includegraphics[angle=0,width=3in]{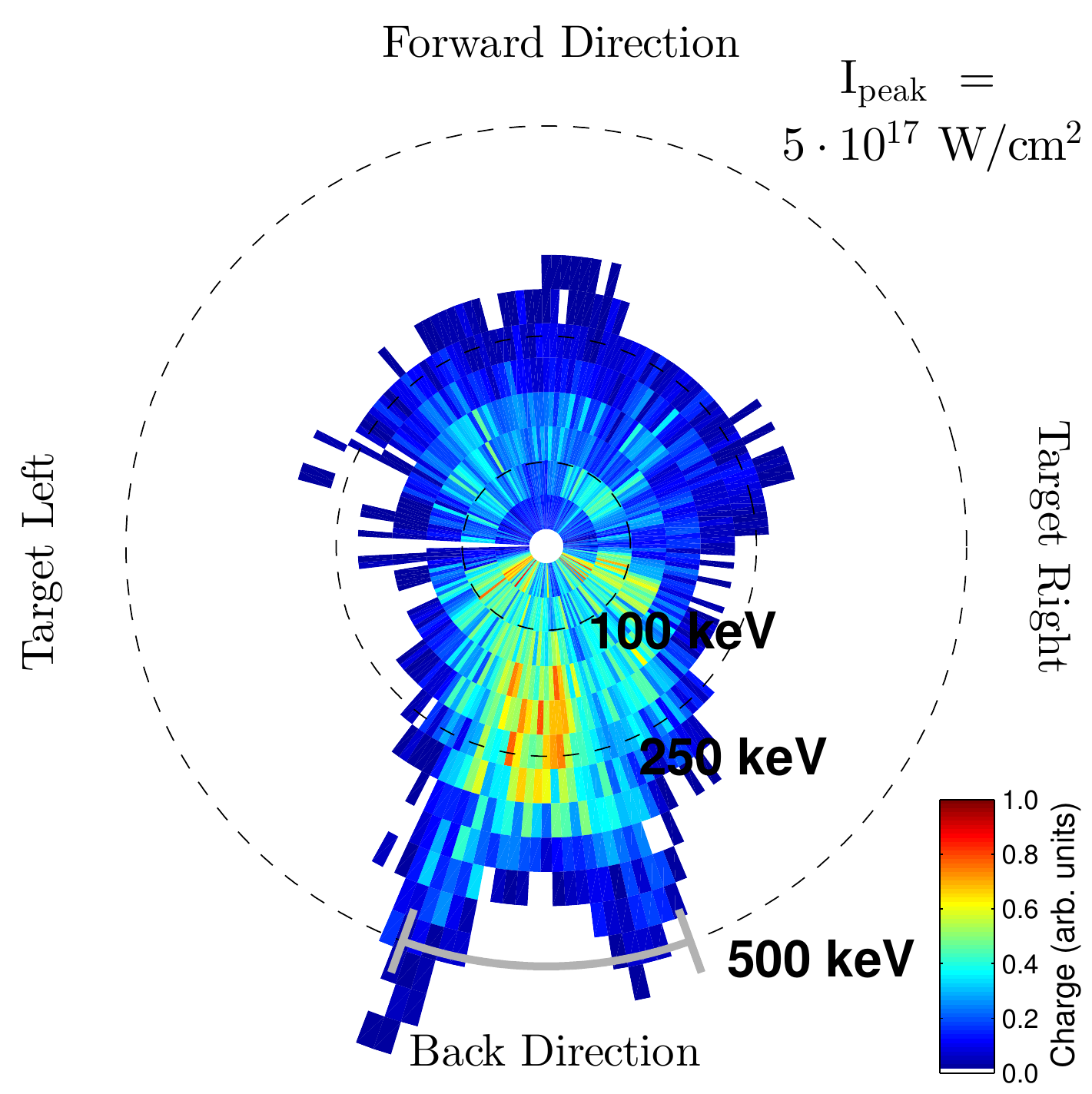}}
\centerline{\includegraphics[angle=0,width=4in]{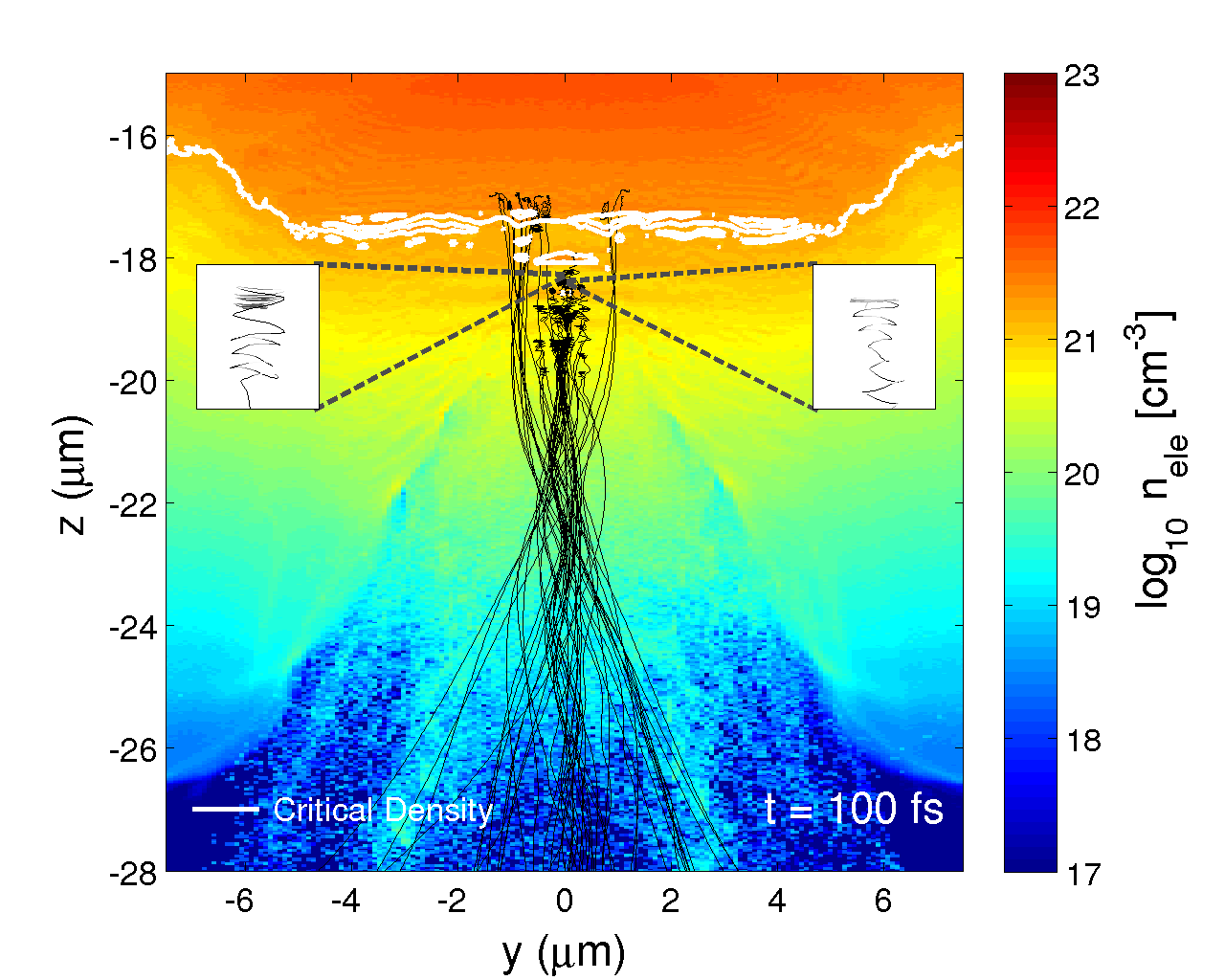} \, \, \includegraphics[angle=0,width=3in]{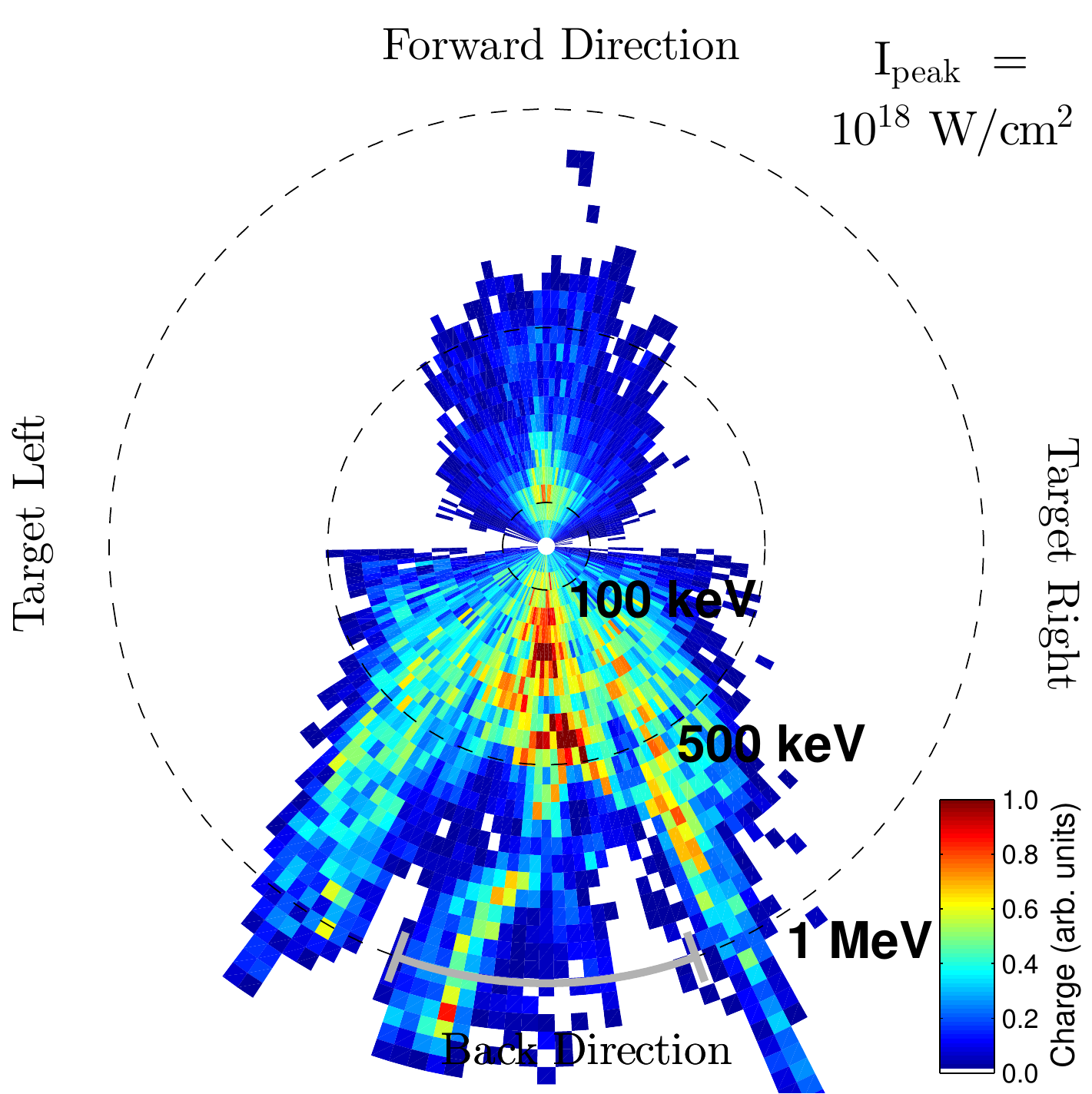}}
\centerline{\includegraphics[angle=0,width=4in]{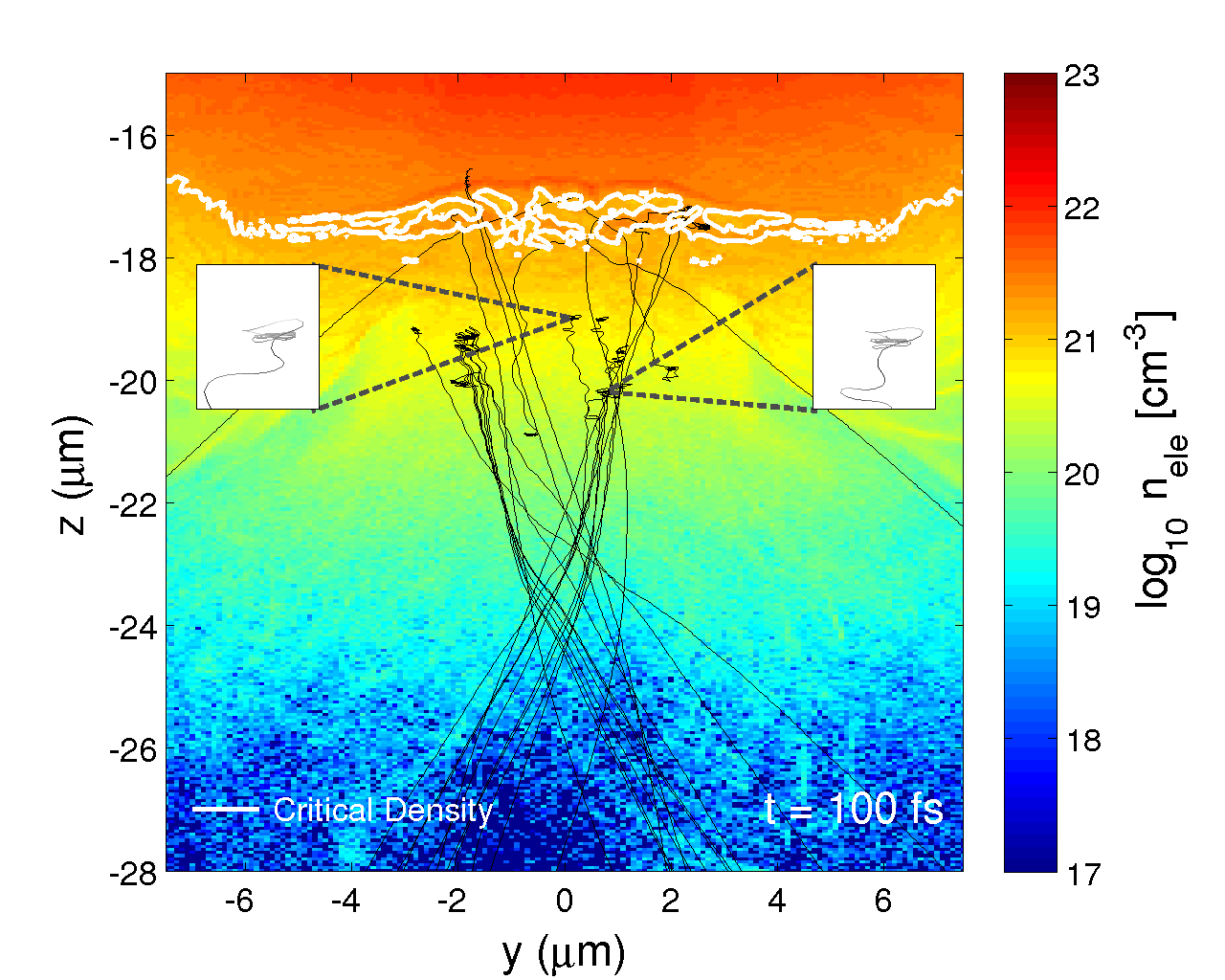} \, \, \includegraphics[angle=0,width=3in]{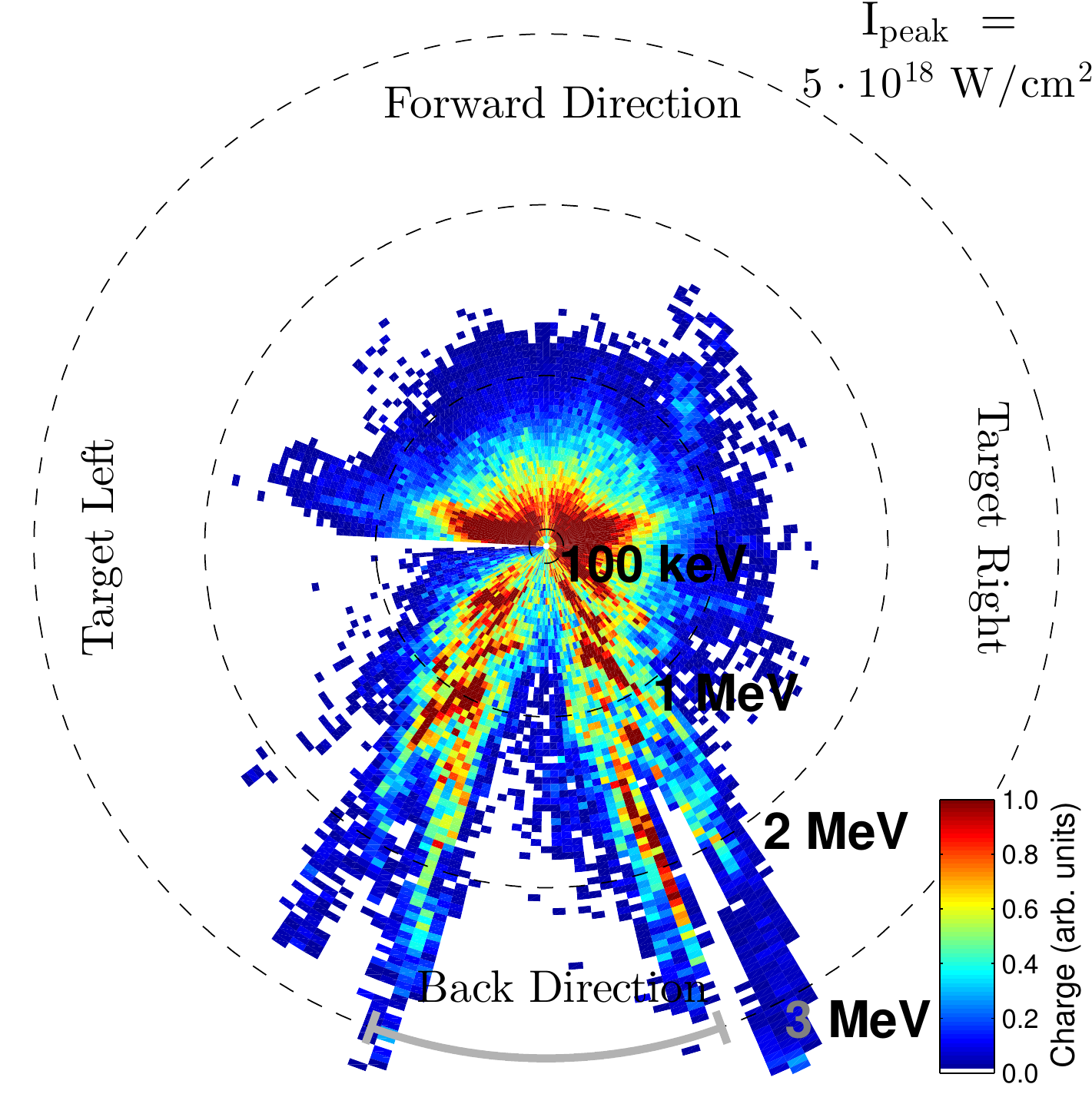}}
\vspace{-0.5cm}
\caption{\emph{Left column}: Electron number density at $t = 100$ fs after the front of the laser pulse reaches the initial critical density surface at $z = -16 \mu$m. Contours of critical electron density ($n_{\rm crit} = 1.72 \cdot 10^{21}$ cm$^{-3}$) are indicated with a thick white line. Also shown are the trajectories of 100 electron macroparticles that are accelerated away from the target. \emph{Right column}: An analysis of energies and escaping angles for electrons that are ejected from the target.}\label{fig:nele}
\end{figure*}

\section{Simulation Results}
\label{sec:results}

Figure~\ref{fig:nele} presents our primary results for the three intensities we consider in detail. The left column highlights the electron number density 100 fs after the front of the laser pulse first arrives at the critical density. The electron trajectories are overplotted with black lines and the position of the critical density contour ($n_{\rm crit} = 1.72 \cdot 10^{21}$~cm$^{-3}$) is also shown with white lines. Within a few microns of the laser axis, this critical density ``surface'' moves forward by about 1.5~$\mu$m in all cases because of further stripping of electrons from the oxygen ions by the laser electric fields. { The right panels also contain zoomed inset figures highlighting two electron trajectories near the laser axis for each intensity. For clarity, the trajectories are displayed with grayscale shading according to $p_z$ momenta, such that motion in the incoming laser direction is white and directly backward motion results in a solid black line.}

The right column of Fig.~\ref{fig:nele} presents analyses of the energies and escaping angles of the electrons that are ejected from the target. These plots were made by recording the charge, energy and angle of every electron macroparticle that left the simulation through the four edges of the simulation grid. The distance from the origin in these plots indicates the kinetic energy of the escaping electrons and the position relative to the origin represents the angle of the escaping electron calculated from the exiting momenta. Positions below the origin represent electrons escaping in specular directions while positions above the origin represent forward-going electrons. A solid gray line shows the convergence angle of laser light on the target in each case. Colors indicate the total charge in each energy-angle bin ($\Delta E = 40$~keV, $\Delta \theta = 2$ deg.).  Except for the highest intensity shown, there are significantly more energetic electrons escaping in the specular direction. The physical mechanisms that eject these electrons will be discussed in the next section.

\begin{figure}
\includegraphics[angle=0,width=3.2in]{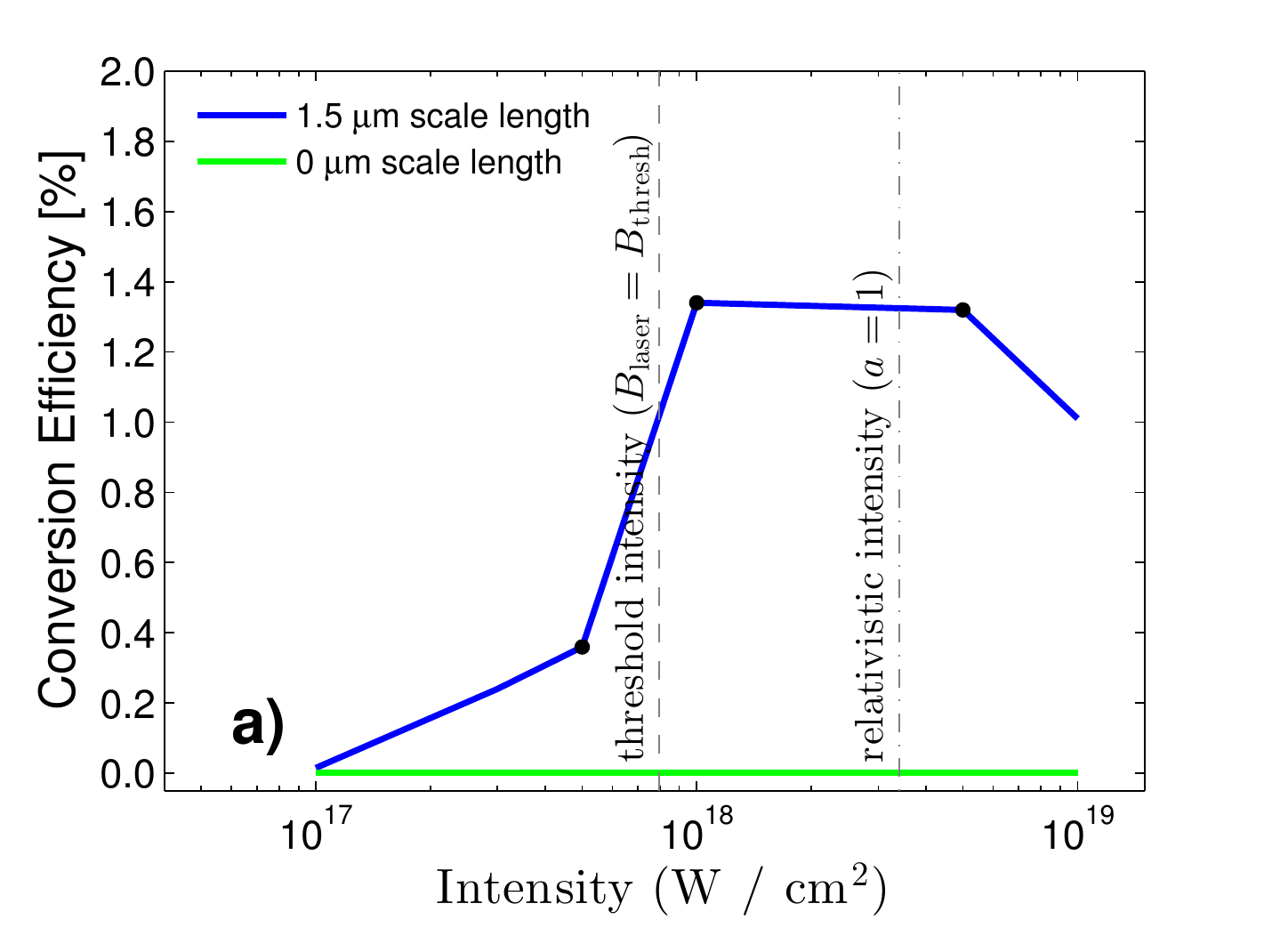}
\includegraphics[angle=0,width=3.2in]{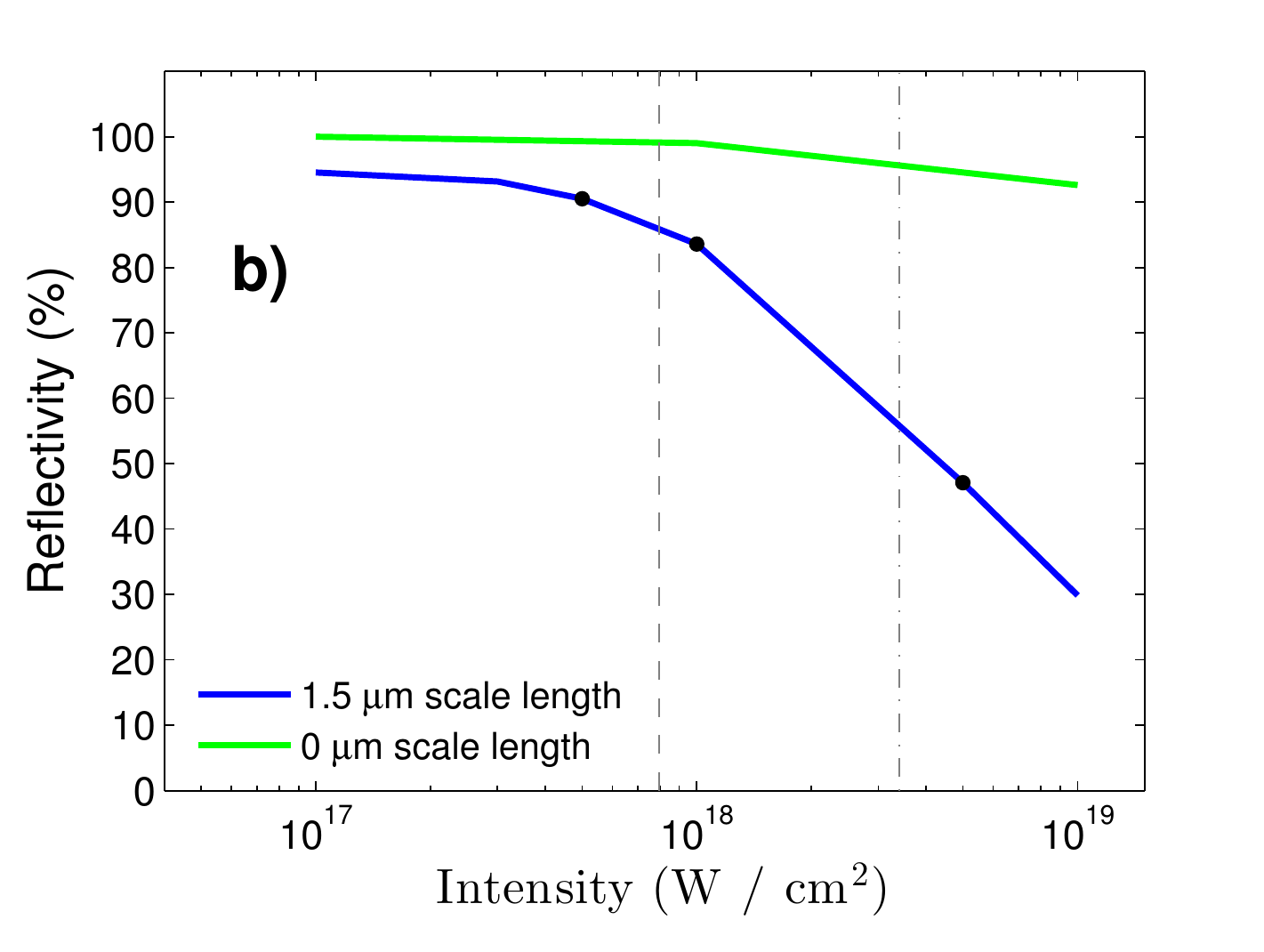}
\vspace{-0.5cm}
\caption{\emph{Panel a}: Conversion efficiency from laser energy to escaping electron energy as a function of intensity. \emph{Panel b}: Measured reflectivity of the target from simulations as a function of intensity. In both panels, results from a $1.5 \mu$m scale length exponential pre-plasma are shown with solid blue lines. Results from a sharp interface (``0 $\mu$m'' scale length pre-plasma) are shown with solid green lines. Black dots show highlighted cases in Fig.~\ref{fig:nele}, which assumed a $1.5 \mu$m scale length pre-plasma. } \label{fig:er}
\end{figure}

Fig.~\ref{fig:er}a shows the laser-to-hot-electron conversion efficiency for specularly directed electrons above 1 keV. We measure this quantity from a number of simulations over wider range of intensities ($10^{17}-10^{19}$~W~cm$^{-2}$) than the three intensities highlighted in Fig.~\ref{fig:nele} which are indicated in Fig.~\ref{fig:er}a with black dots. Our choice to highlight the intensities $5 \cdot 10^{17}$, $10^{18}$ and $5 \cdot 10^{18}$~W~cm$^{-2}$ was informed by Fig.~\ref{fig:er}a. At $5 \cdot 10^{17}$~W~cm$^{-2}$ the conversion efficiency is rising with intensity, at $10^{18}$~W~cm$^{-2}$ the conversion efficiency reaches 1.3\%, and $5 \cdot 10^{18}$~W~cm$^{-2}$ is a high-conversion efficiency case with a relativistic intensity.

Also shown in Fig.~\ref{fig:er}a, is the conversion efficiency for targets \emph{without} a pre-plasma, which is equivalent to a ``0 $\mu$m'' scale length. For a wide range of intensities there are essentially no specularly-accelerated electrons ejected off the target. Instead the target, which features a very sharp interface, simply reflects the incident laser light. This result from simulation qualitatively matches the empirical result from the experiment. The formation of a specularly-accelerated electron beam {\it requires} the presence of pre-plasma \cite{Morrison_etal2015}. { Many other experiments with near-solid density targets have demonstrated the importance of the pre-plasma (e.g. \cite{Ovchinnikov_etal2011}) , but relatively few experiments with such targets have noticed a benefit from the presence of a pre-plasma (for exceptions, c.f., \cite{Dudnikova_etal2003,Esirkepov_etal2014}).}

Fig.~\ref{fig:er}b highlights the reflectivity of the target as a function of intensity. This was calculated by integrating the electromagnetic field energy that leaves the simulated target and comparing to the incident laser energy. The reflectivity was overall quite high with $\sim 80-90\%$ reflectivity for intensities up to $10^{18}$ W cm$^{-2}$ for the $1.5 \mu$m scale length results and even higher ($\geq 90\%$) for the sharp interface. These numbers are experimentally reasonable. \citet{Panasenko_etal2010} report reflectivities of $\approx 70\%$ from short-pulse laser interactions with a water film target at $10^{16}$ W cm$^{-2}$ and the trend in their Fig.~3 suggests that the reflectivity would peak above $70 \%$ at some intensity above $10^{16}$ W cm$^{-2}$.

At the highest intensities in Fig.~\ref{fig:er}b, the reflectivity drops as many other studies have observed. This trend is well understood as a consequence of relativistic absorption (e.g. \citet{Levy_etal2013} and references therein). The $\approx50\%$ reflectivity for the $5 \cdot 10^{18}$ W cm$^{-2}$ case in Fig.~\ref{fig:er}b provides some explanation for the significant numbers of forward-going electrons observed in the bottom right panel of Fig.~\ref{fig:nele}. While appreciable numbers of electrons are accelerated by the reflected laser pulse in this case, many electrons in the pre-plasma will instead be accelerated into the target in response to the overall forward-going laser fields.

\begin{figure}[t]
\includegraphics[angle=0,width=3.7in]{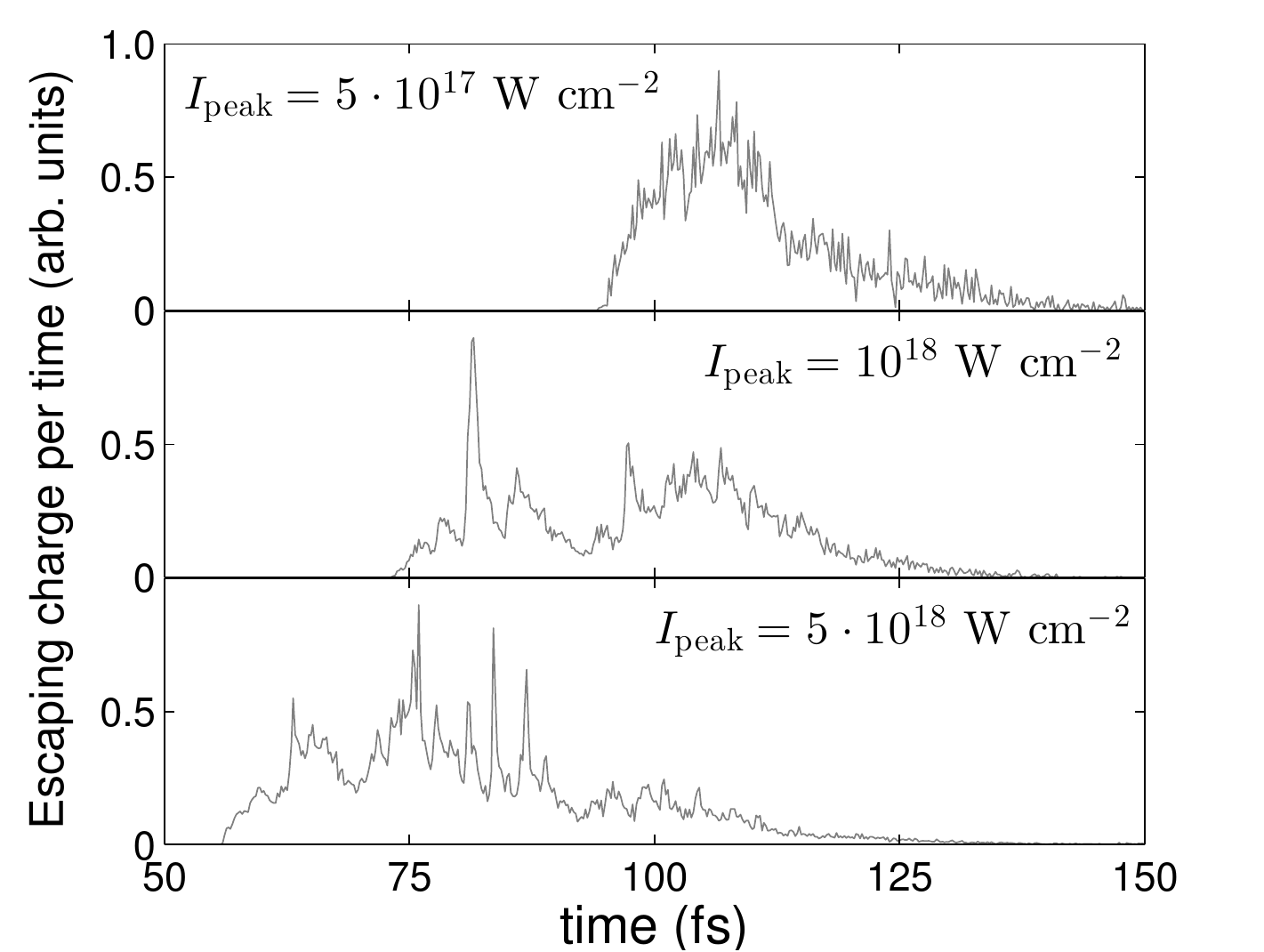}
\vspace{-0.7cm}
\caption{A plot of the electron charge leaving the PIC simulations through the specular boundary ($z = -30 \mu$m) as a function of time (0.2 fs bins). For each intensity, electron macroparticles that escaped with kinetic energies $> 100$~keV were recorded. Results with different kinetic energy thresholds are qualitatively similar.} \label{fig:qleaving}
\end{figure}

The last empirical observation from the simulations worthy of note is that many of the escaping electrons leave the target in sub-fs bunches. Fig.~\ref{fig:qleaving} shows the amount of charge per time leaving the edge of the simulation at $z = -30~\mu$m binned in increments of 0.2~fs (i.e. 4$\times$ the timestep). Particularly apparent at $10^{18}$~W~cm$^{-2}$ and $5 \cdot 10^{18}$~W~cm$^{-2}$ intensities are moments where a substantial amount of charge leaves the edge of the simulation in under a femtosecond. There may be sub-fs bunching in the $5\cdot 10^{17}$ W cm$^{-2}$ case as well. Similarly short bunches of electrons have been observed in simulations of solid density target irradiation by \citet{Naumova_etal2004} who emphasize the novelty of using these bunches to create secondary light sources with ultra-short or attosecond features. From the standpoint of the AFRL experiment, confirmation of the bunched nature of the escaping electrons (e.g. through detection of coherent transition radiation \cite{Zheng_etal2003}) remains an important goal for future work.

\section{Mechanisms of Electron Acceleration} 
\label{sec:mech}

\subsection{Outline}

We have come to the following conclusions regarding the precise mechanisms of electron acceleration in the simulations:

\begin{enumerate}

\item 
Electrons are launched both towards and away from the target through interactions with the standing wave created by the overlap of forward and reflected light. { Electrons can be launched if they are positioned near the half-way point between a node and an anti-node of the electric field. These locations experience both strong electric and magnetic fields due the standing wave.}

\item
Electrons that are launched from the standing wave in the specular (back reflected) direction have an opportunity to get an extra ``boost'' in kinetic energy by interacting with the reflected laser pulse.

\item
This kinetic energy ``boost'' from the reflected laser pulse is more effective than expected from simple assumptions. Further investigation reveals that the distorted nature of the reflected pulse helps to collimate and amplify the energies of the escaping electrons. 

\item
Quasi-static electric fields arising from the modification of the electron density profile by the standing wave can also play a role in increasing the electron kinetic energies.

\end{enumerate}

Conclusions 1 \& 2 will be discussed in \S~\ref{sec:sw}. Conclusions 3 \& 4 will be explained in \S~\ref{sec:real}. These conclusions come from analysis of both the realistic PIC simulations described in the previous section and a set of additional ``idealized'' PIC simulations that will be described in \S~\ref{sec:ideal}. The ``idealized'' simulations are designed to isolate and study the standing wave acceleration mechanism using the same particle tracking tools used in the realistic simulations.

\subsection{Standing Wave Acceleration}
\label{sec:sw}

\begin{figure*}
\includegraphics[angle=0,width=3.5in]{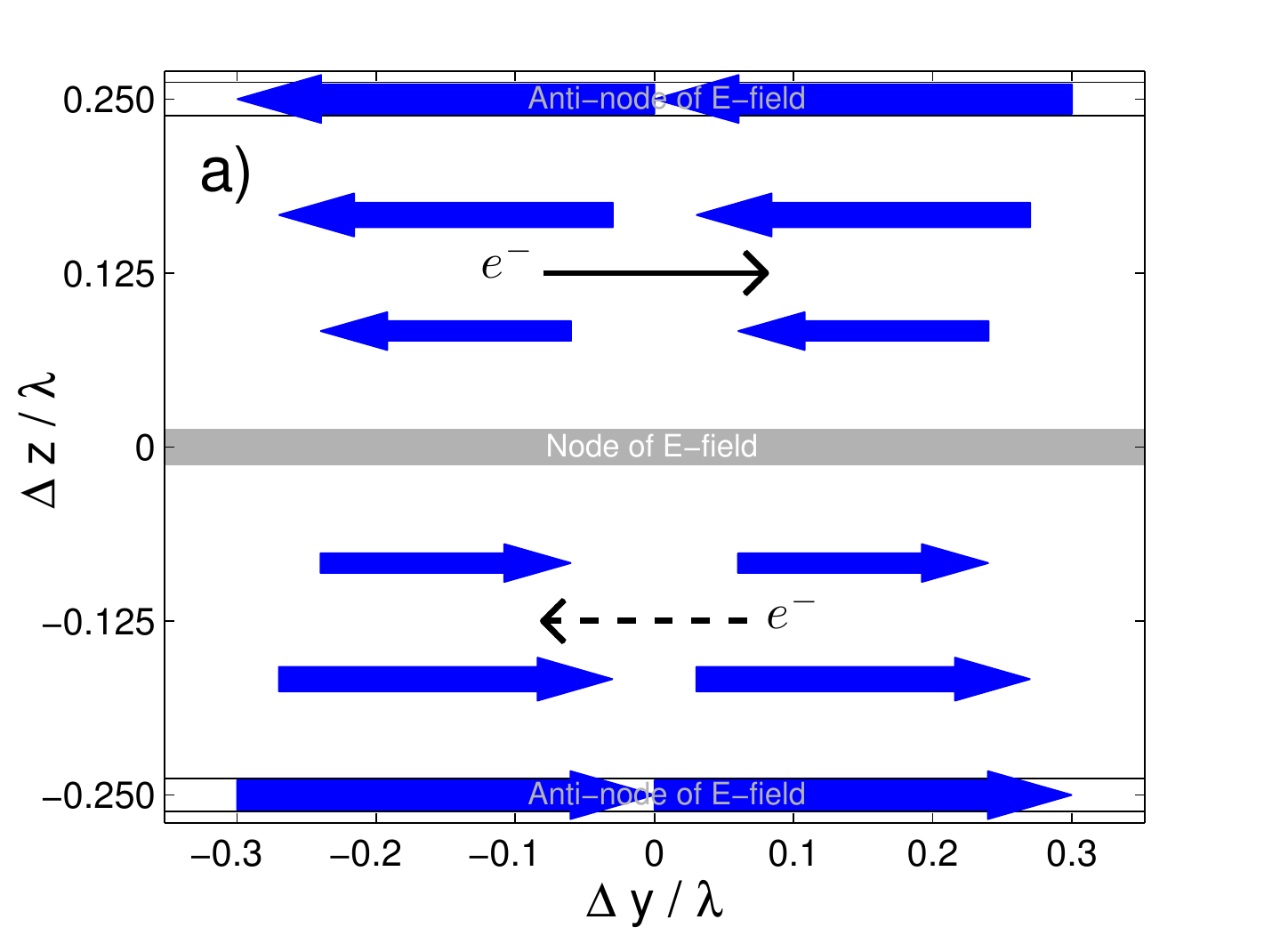}\includegraphics[angle=0,width=3.5in]{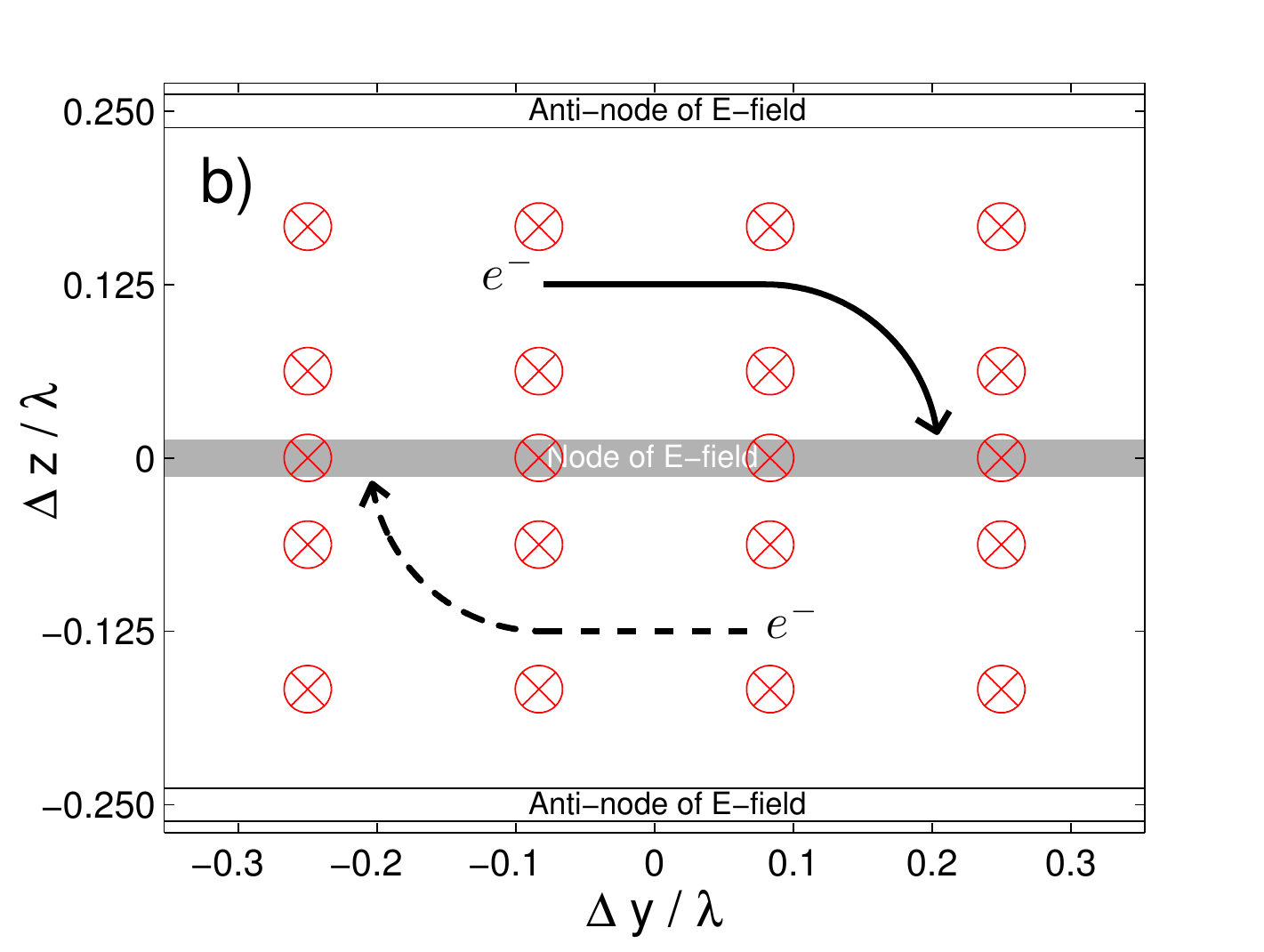}
\includegraphics[angle=0,width=3.5in]{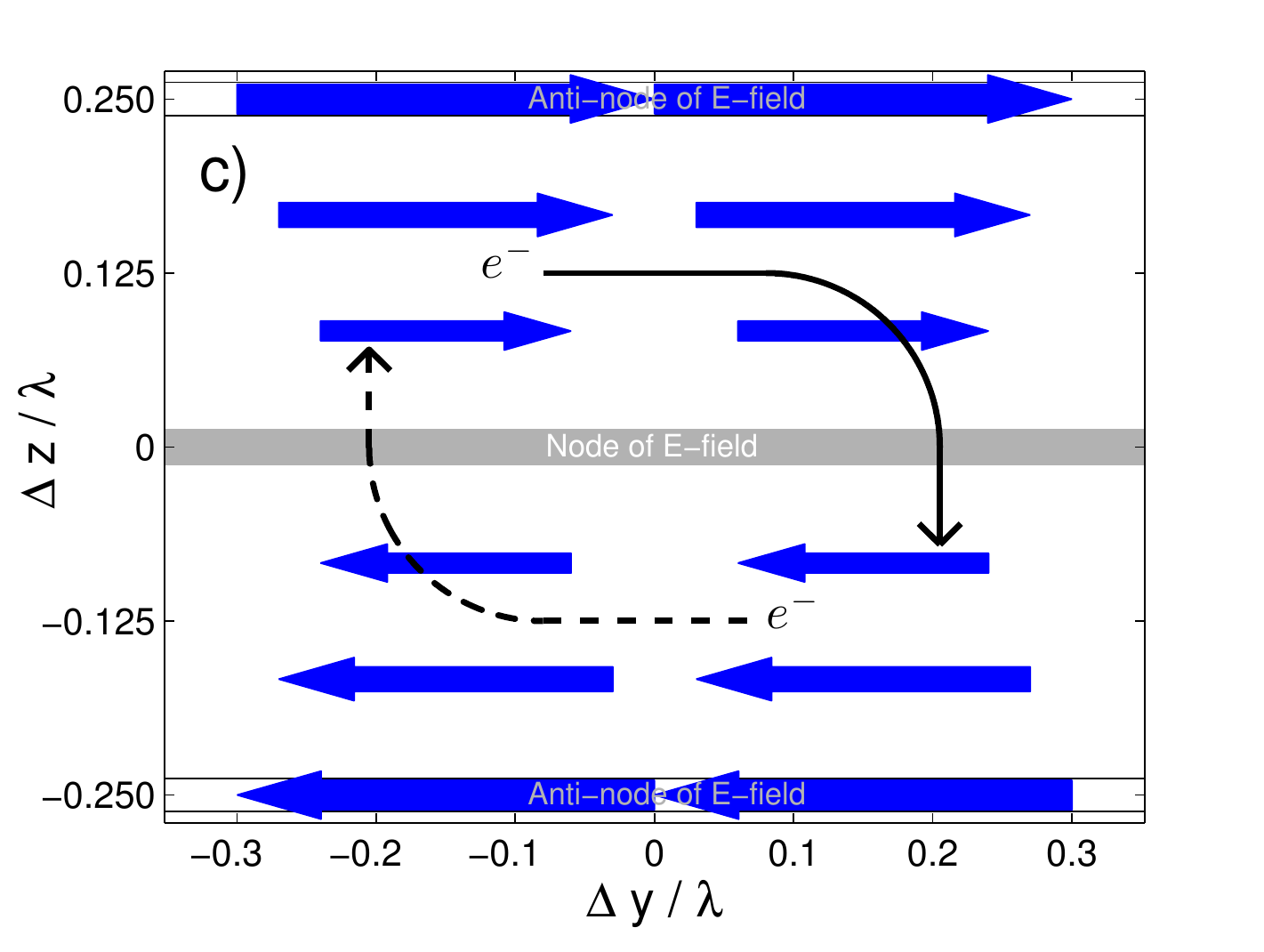}\includegraphics[angle=0,width=3.5in]{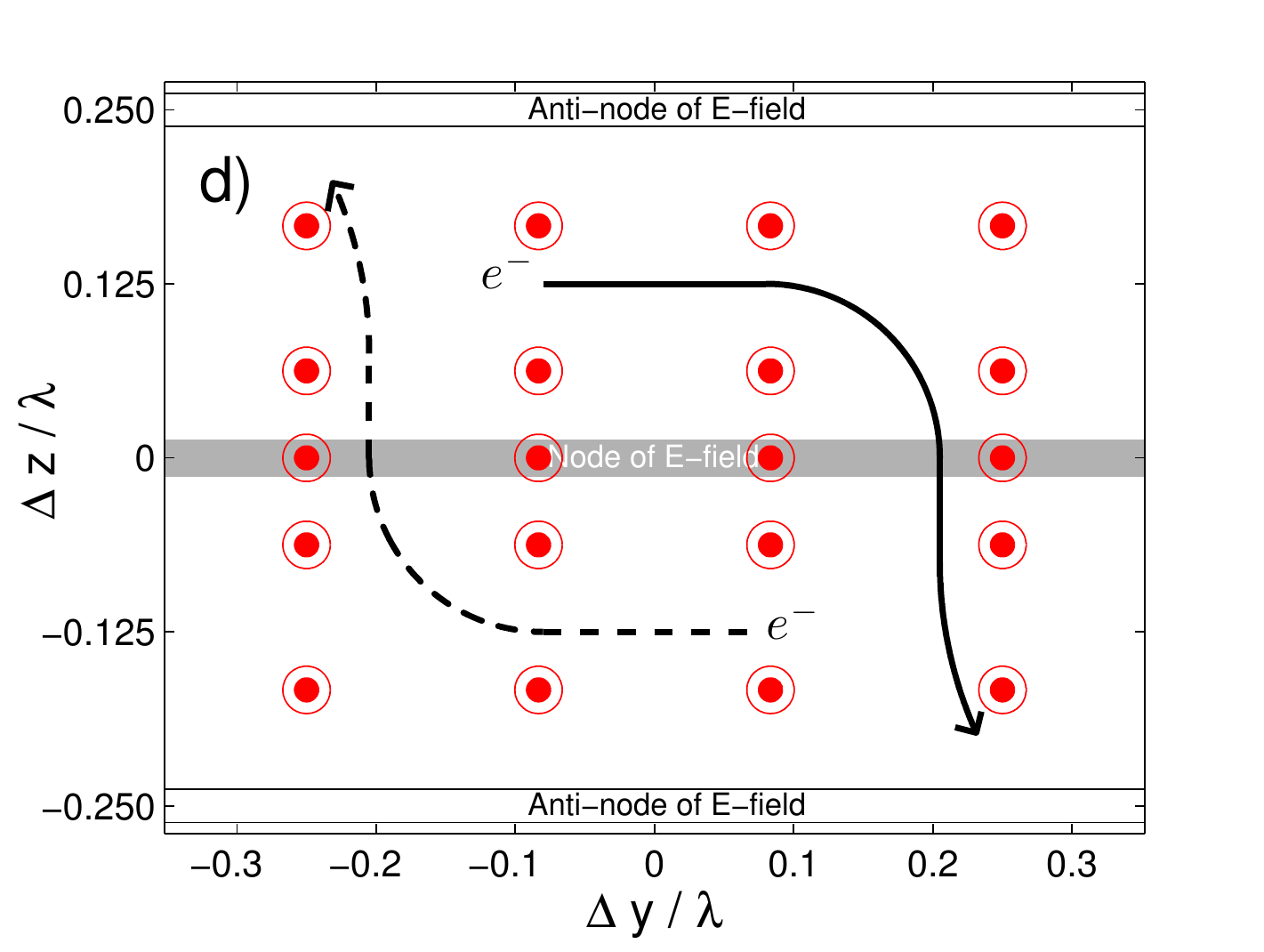}
\vspace{-0.3cm}
\caption{An illustration of standing wave acceleration at relativistic intensities ($a_0 \gtrsim 1$) in its four different stages. The target is located at some $+z$ value above the area shown. Thick solid lines indicate electrons that are ultimately accelerated away from the target while dashed lines indicate electron trajectories accelerated into the target. Panel a. illustrates the ``push'' phase, Panel b. illustrates the ``rotate'' phase, and Panel c. illustrates the ``drift'' phase. Finally, in Panel d., if the magnetic fields are substantially weaker than during the ``rotate'' phase the electron will experience a mild deflection and continue its overall motion. }\label{fig:sw}
\end{figure*}

Before describing the idealized simulations, we provide some context in this subsection and describe some simple estimates for the electron kinetic energies. \citet{Kemp_etal2009} identified standing wave acceleration as an important mechanism for accelerating electrons in the \emph{forward direction} for normal incidence and at intensities $\sim 10^{20}$~W~cm$^{-2}$. This mechanism is an efficient accelerator of electrons when the overlap of the forward-going and reflected laser pulse gives rise to a standing wave pattern, i.e., 
\begin{eqnarray}
E_y (y = 0, z, t) = 2 \, E_{y0} \, \sin \left( \frac{2 \pi}{\lambda} (z - z_{\rm c}) \right) \sin( \omega t) \label{eq:swE}\\
B_x (y = 0, z, t) = 2 \, B_{x0} \, \cos \left( \frac{2 \pi}{\lambda} (z - z_{\rm c}) \right) \cos( \omega t) \label{eq:swB}
\end{eqnarray}
where $z_c$ is the position of a sharp interface where the reflection occurs and $\omega$ is the angular frequency of the laser. \citet{Kemp_etal2009} consider electron acceleration in these time and space-varying electric and magnetic fields for highly-relativistic electrons where $v \sim c$. Fig.~\ref{fig:sw} illustrates the acceleration of electrons in both the forward-going (dashed lines) and specular directions (solid lines). As in other figures, the $-z$ direction is the direction away from the target. Note that because of the constructive and destructive interference, there are moments in every laser cycle where $E_y$ is zero and $|B_x|$ is peaked, and moments where $B_x$ is zero and $|E_y|$ is peaked (Eqs.~\ref{eq:swE} \& \ref{eq:swB}). Fig.~\ref{fig:sw} highlights these times, which are crucial for understanding the acceleration of the electrons.

In their paper, \citet{Kemp_etal2009} numerically integrated the trajectories of electrons in a simple standing wave (Eqs.~\ref{eq:swE} \& \ref{eq:swB}). \cite{Kemp_etal2009} concluded that the maximum momenta attainable by electrons is given by
\begin{equation}
p_{\rm max} = 1.45 \, a_0  \label{eq:pmax}
\end{equation}
where $p_{\rm max}$ is a normalized to $mc$. \citet{Kemp_etal2004} presented evidence from 1D(3$v$) simulations (their Fig. 4) that $p_{\rm max}$  is a reasonably accurate estimate for a cutoff feature in their forward-going energy distribution. As illustrated in our Fig.~\ref{fig:sw}, Eq.~\ref{eq:pmax} should apply equally well to electrons accelerated away from the target. However, in this case, once outside of the standing wave, electrons should gain additional momenta from interacting with the reflected pulse. 

\begin{figure}
\centerline{\includegraphics[angle=0,width=3.25in]{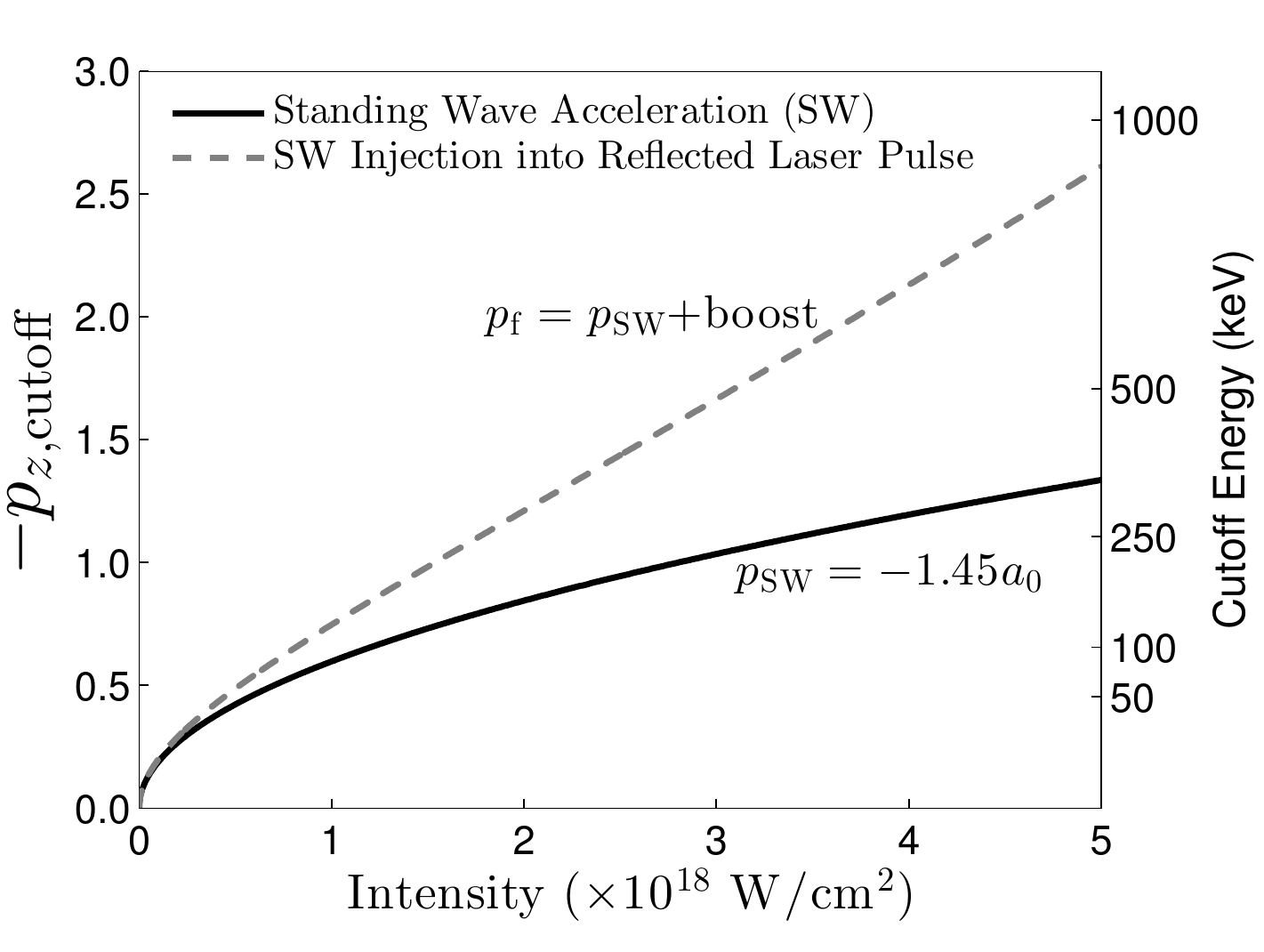}}
\vspace{-0.4cm}
\caption{Simple estimates for the cutoff momentum (and corresponding kinetic energy). The solid black line shows the 1.45 $a_0$ cutoff from \citet{Kemp_etal2009}, who considered standing wave acceleration in the forward direction. The gray dashed line shows the prediction of Eq.~\ref{eq:boosted} which is a simple plane-wave estimate for the momentum boost from electrons launched by the standing wave into the reflected laser pulse. This estimate is compared to results from idealized simulations in Fig.~\ref{fig:ideal}.} \label{fig:boost}
\end{figure}

A simple estimate for this additional momentum comes from \cite{Yu_etal2000} by applying a plane-wave approximation \cite{LandauLifshitz} to electrons moving with non-zero momenta away from the target. These considerations yield
\begin{equation}
p_{z{\rm f}} = p_{z0} + \frac{-a_0^2}{2(p_{z0} + \sqrt{1+p_{z0}^2})} \label{eq:boosted}
\end{equation}
for the ``final'' momentum, $p_{z{\rm f}}$, of the electron from interacting with the reflected laser light where $p_{z0}$ is the initial momentum (e.g. provided by the standing wave mechanism), and $a_0$ is the $a$-value of the laser field. Note that since motion away from the target implies $p_{z0} < 0$, as $-p_{z0}$ becomes large the denominator of the second term in Eq.~\ref{eq:boosted} tends to zero and the additional momentum provided by this second term becomes significant. Fig.~\ref{fig:boost} uses Eq.~\ref{eq:boosted} to show how the cutoff momentum, and corresponding cutoff energy, scale with intensity by assuming $p_{z0} = -1.45 \, a_0$. This expectation for the cutoff energy will be compared with the results of ``idealized'' 2D(3$v$) PIC simulations discussed in the next subsection.


\begin{figure*}
\centerline{\includegraphics[angle=0,width=3in]{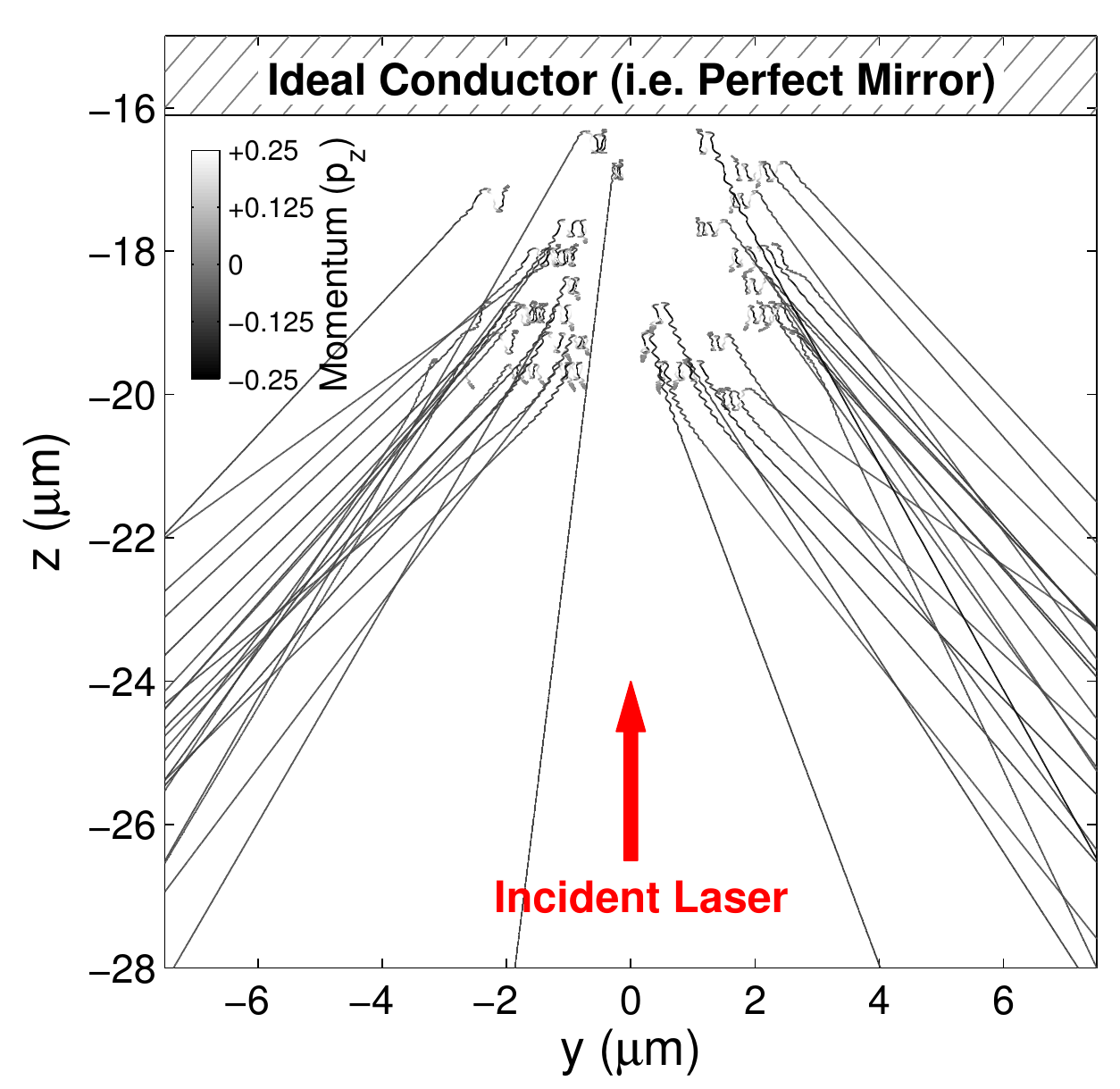} \, \, \, \, \includegraphics[angle=0,width=3in]{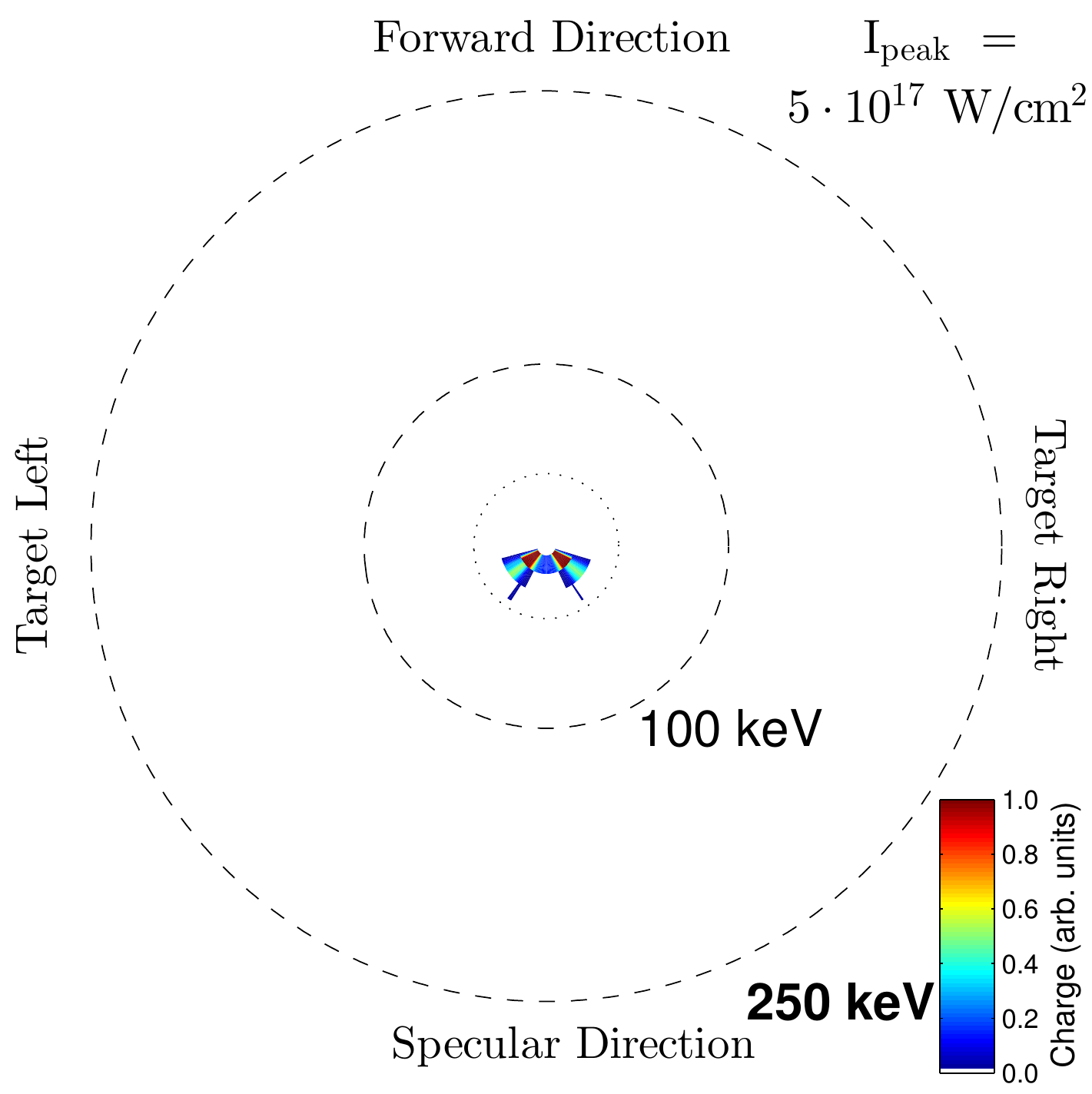}}
\centerline{\includegraphics[angle=0,width=3in]{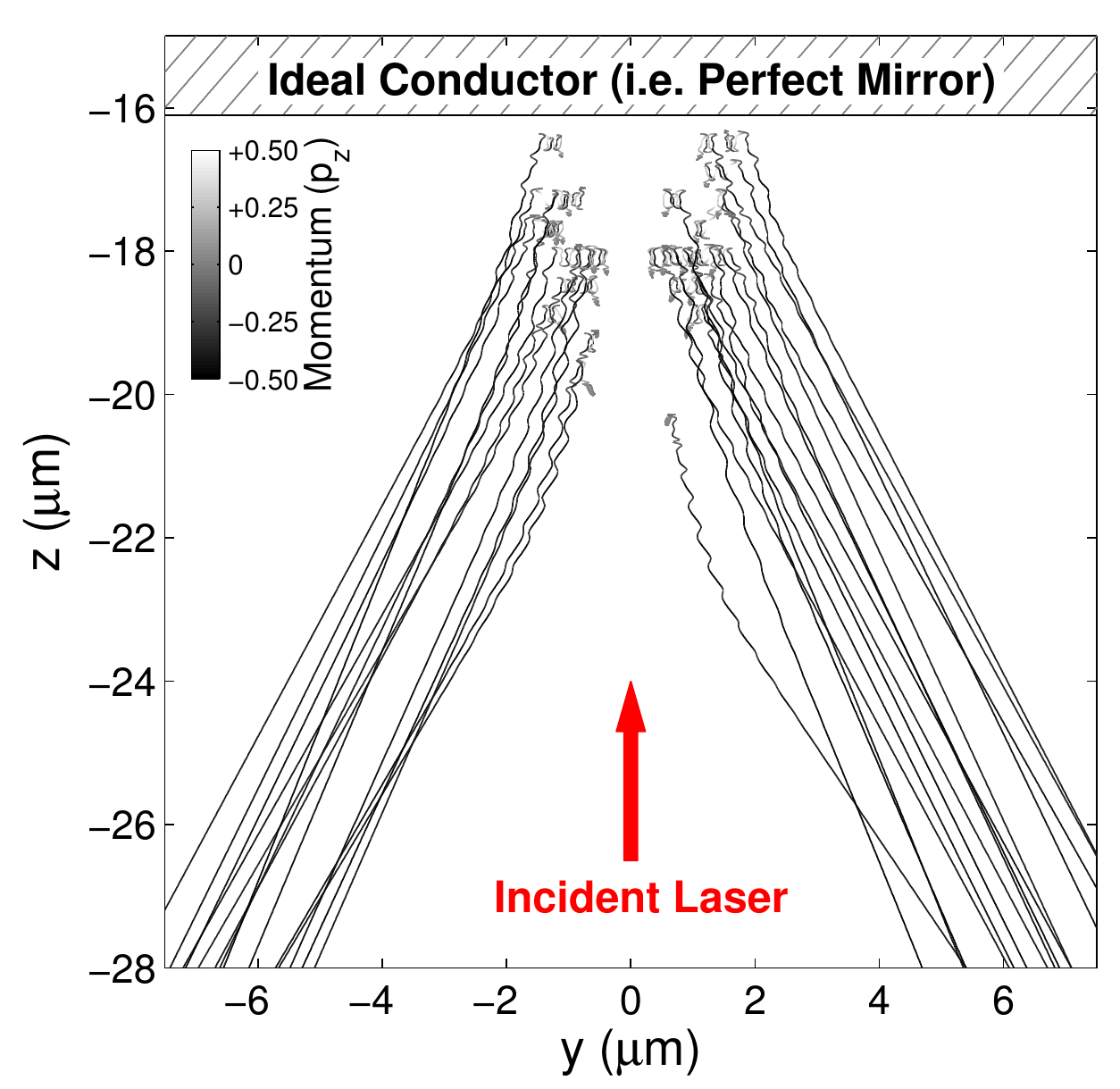} \, \, \includegraphics[angle=0,width=3in]{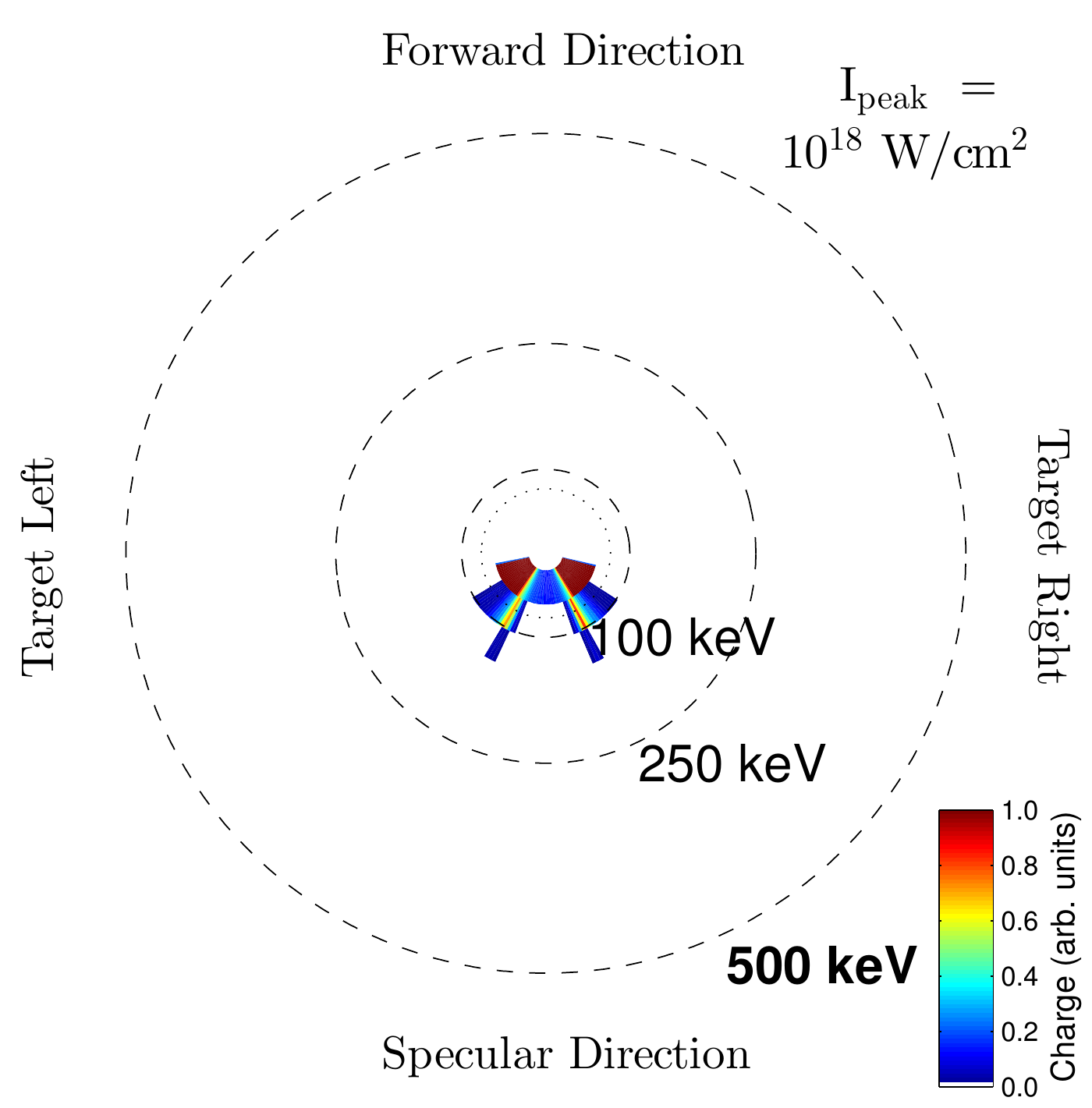}}
\centerline{\includegraphics[angle=0,width=3in]{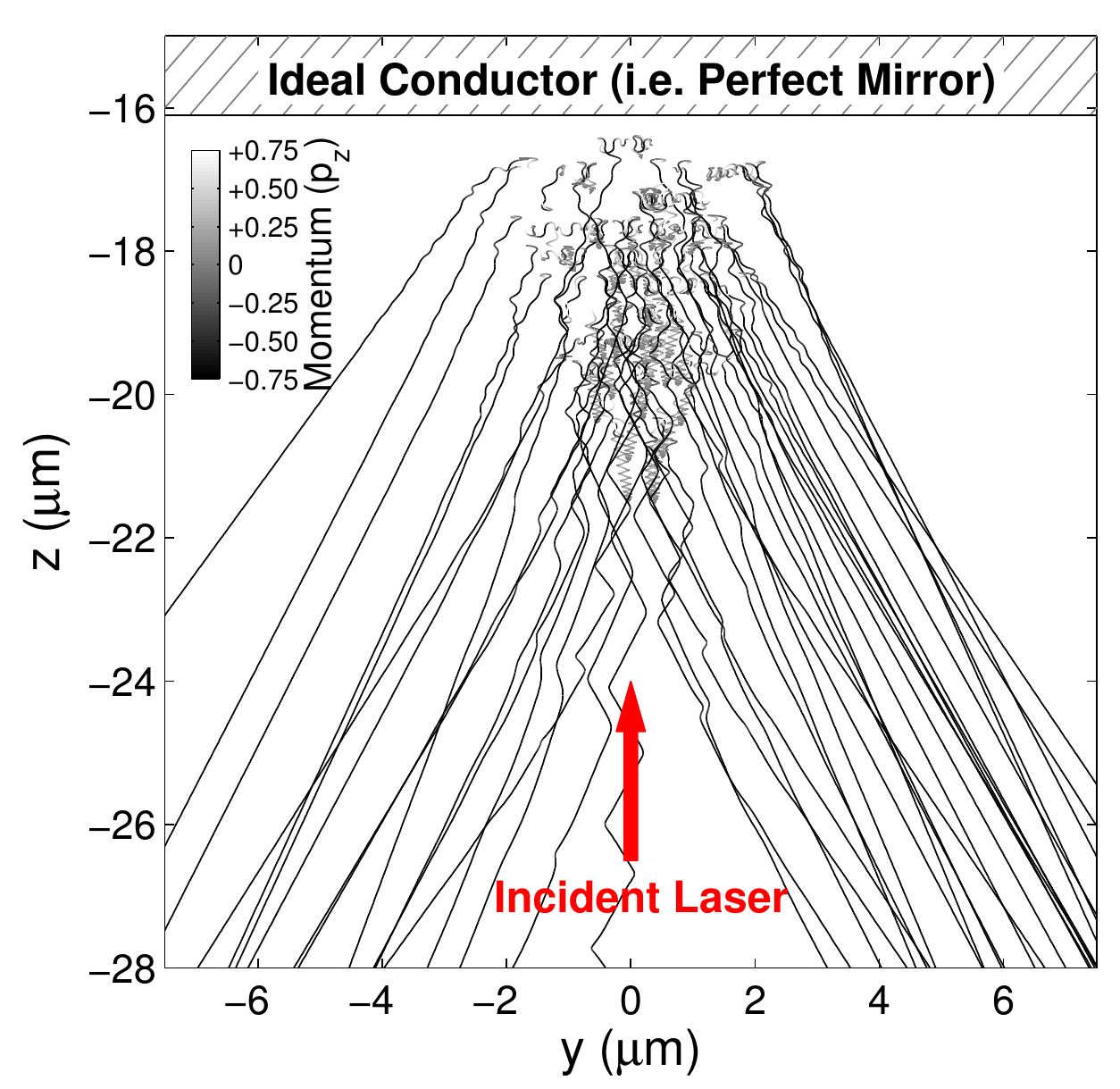} \, \, \includegraphics[angle=0,width=3in]{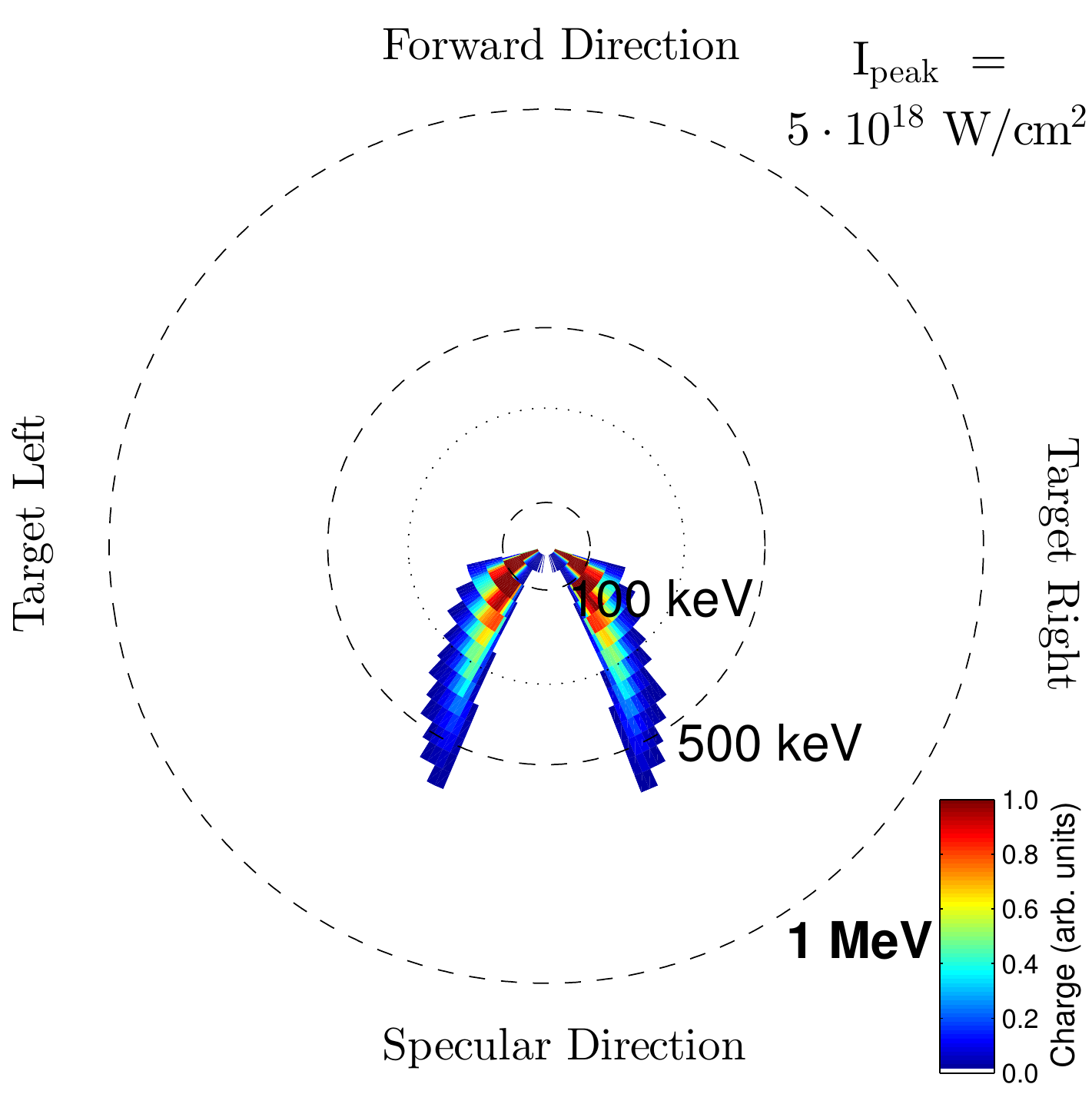}}
\vspace{-0.4cm}
\caption{\emph{Left column}: Electron trajectories from idealized simulations where the same ultra-intense laser pulses are incident on a perfectly reflecting interface with an extremely diffuse pre-plasma. Trajectories have been color coded according to their $p_z$ momenta towards (gray or white) or away (dark gray or black) from the target. Although the trajectories are highly chaotic in each case, this shading helps to highlight the moment when the macroparticle is flung away from the target. \emph{Right column}: An analysis of angles and kinetic energies of electrons that are accelerated away from the target (compare with right column of Fig.~\ref{fig:nele}). The dotted circle represents the kinetic energy cutoff predicted by Eq.~\ref{eq:boosted}.}\label{fig:ideal}
\end{figure*}

\subsection{Idealized PIC Simulations}
\label{sec:ideal}

Idealized 2D(3$v$) PIC simulations were performed with the same intensities highlighted in \S~\ref{sec:sims}, including the same spot size and temporal profile in \S~\ref{sec:sims}, but with a different target geometry. The idealized aspect of these simulations stems from using an extremely low pre-plasma density ($10^{10}$ cm$^{-3}$) with a flat density profile\footnote{The pre-plasma density is set low enough that charge separation effects should be minimal. As such the electron macroparticles should respond to the laser electric and magnetic fields as tracer particles. The ions in the simulation were immobile and fixed in ionization state in order to create a neutralizing background. The results we present in this section are insensitive to the exact value of this extremely low density as one would expect.} and by using an ideal conductor, which acts as a perfect mirror, to reflect the laser pulse. With this choice the electric and magnetic fields along the laser axis and where the forward and reflected pulses overlap are well described with by simple standing wave (Eqs.~\ref{eq:swE} \& \ref{eq:swB}). In the idealized simulations the conductor was placed at $z_{\rm c} = -16~\mu$m, which is approximately where the critical density appears in the realistic simulations described in \S~\ref{sec:sims}.

Fig.~\ref{fig:ideal} presents the main results from these idealized simulations. Much like Fig.~\ref{fig:nele}, the left column shows the electron trajectories while the right column shows an analysis of the energies and angles of the escaping electrons. The electron trajectories in the left column plots have been color coded according to their momenta towards or away from the target. Although the trajectories are highly chaotic in each case, this shading helps to highlight the moment when an electron is flung away from the target. Once ``launched'' in the -$z$ direction, an electron may be deflected and accelerated by the reflected laser pulse but it will generally continue moving away from the target until it exits the simulation. { The $5 \cdot 10^{18}$~W~cm$^{-2}$ results in Fig.~\ref{fig:ideal} naturally provide the best illustration for standing wave acceleration with $a_0 \gtrsim 1$. A close look at the trajectories near the laser axis and near $z \sim -17$~$\mu$m do exhibit the quarter-circle turn away from the target as described in Fig.~\ref{fig:sw}. Other trajectories show motion in the $x$-direction followed by a somewhat more than 90-degree turns away from the target that still produce electron motion away from the target with significant $p_z$ momentum. These trajectories will be discussed in \S~\ref{sec:mod} which examines the nature of standing wave acceleration for $a_0 \sim 0.5$. The trajectories from the realistic simulations, highlighted earlier in Fig.~\ref{fig:nele}, show a mixture of these behaviors and other chaotic motions that will be discussed in \S~\ref{sec:real} \& \S~\ref{sec:mod}.}

Eq.~\ref{eq:boosted} does a reasonable job of predicting the cutoff energy in the idealized simulations, as seen in the right-hand column of Fig.~\ref{fig:ideal}. The dotted circles show the kinetic energy corresponding to the momentum cutoff of Eq.~\ref{eq:boosted}. This prediction matches well the $10^{18}$~W~cm$^{-2}$ and $5 \cdot 10^{18}$~W~cm$^{-2}$ results where the dotted circle appears at energies just beyond the red-shaded zones that indicate the energies and angles where most of the electrons exit the simulation. The dotted circle substantially overpredicts the analogous cutoff in the $5 \cdot 10^{17}$~W~cm$^{-2}$ case. This can be attributed to the inaccuracy of using $p_{z0} = -1.45 \, a_0$ for $a_0 \sim 0.5$ (c.f. Table~\ref{tab:intensities}). As mentioned earlier, this relation only applies when the electron velocity is close to the speed of light \cite{Kemp_etal2009}. Standing wave acceleration for $a_0 \sim 0.5$ will be discussed in \S~\ref{sec:mod}.

\subsection{Comparison to the realistic case}
\label{sec:real}

Comparing the angle and energy analysis shown on the right-hand column of Fig.~\ref{fig:ideal} to the right-hand column of Fig.~\ref{fig:nele}, there are a number of striking differences. Most prominently, the electron kinetic energies in the realistic case are considerably more energetic than the idealized case for each intensity. The difference is so striking that the energy scale needed to be expanded in Fig.~\ref{fig:ideal} in order to adequately show the angular distributions of the escaping electrons. 

There are also key differences in the exiting angles of the escaping electrons. For each intensity shown for the idealized case, Fig.~\ref{fig:ideal} indicates that there are two preferred angles and relatively few electrons escape along the laser axis. The right-hand column of Fig.~\ref{fig:nele}, shows that intensities of $10^{18}$~W~cm$^{-2}$ and $5~\cdot~10^{18}$~W~cm$^{-2}$ have preferred angles of escape away from the laser axis, not unlike the ideal case. However, both $5 \cdot 10^{17}$~W~cm$^{-2}$ and $10^{18}$~W~cm$^{-2}$ indicate significant electrons escaping parallel to the laser axis.

\begin{figure*}
\centerline{\includegraphics[angle=0,width=2.84in]{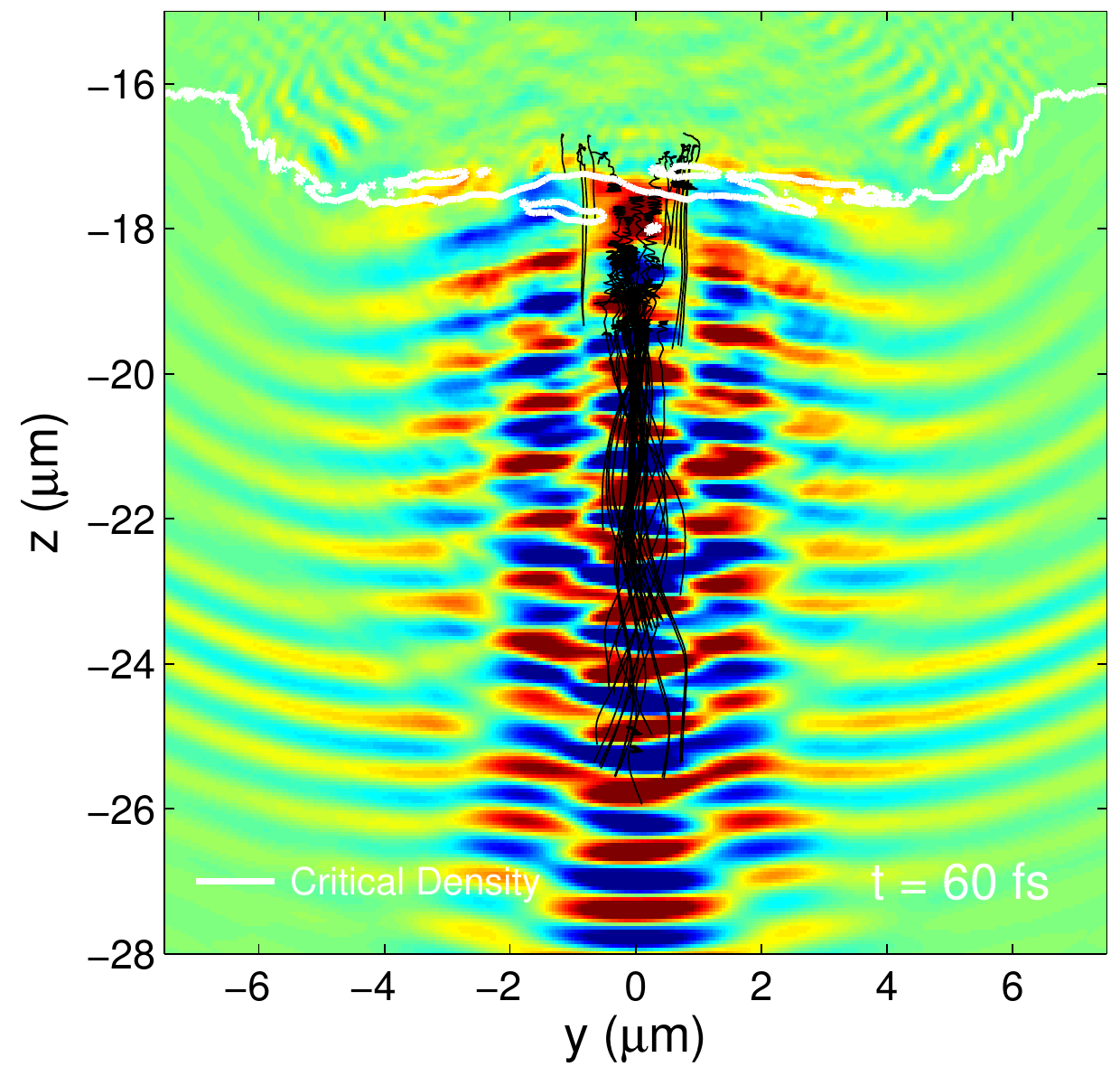}\includegraphics[angle=0,width=3in]{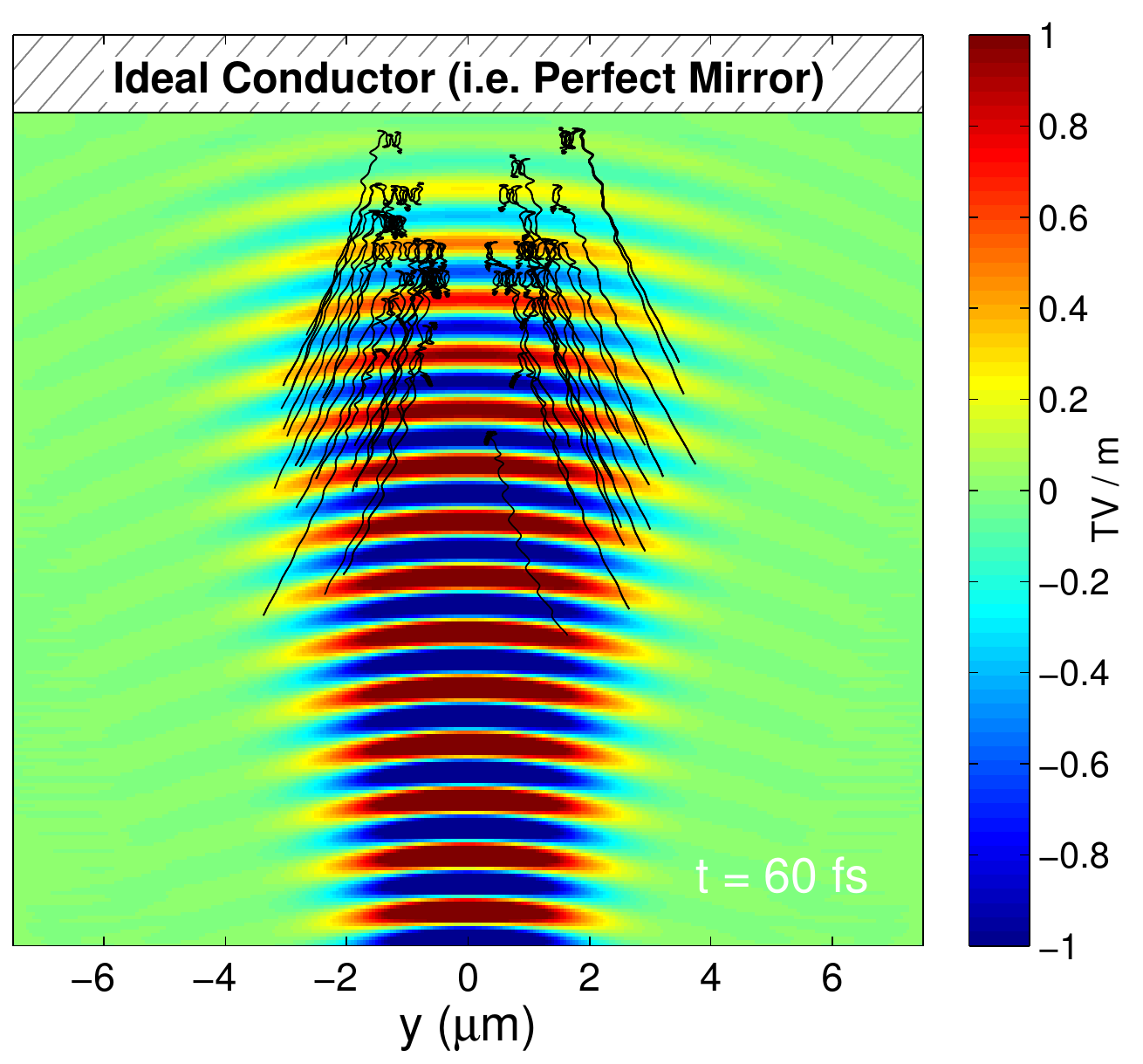}}
\vspace{-0.4cm}
\caption{ \emph{Left panel}: Transverse electric fields and electron trajectories from the $10^{18}$ W cm$^{-2}$ simulation with a realistic $1.5 \mu$m scale length pre-plasma highlighted in \S~\ref{sec:results}. \emph{Right panel}: Transverse electric fields and electron trajectories from an ``idealized'' $10^{18}$ W cm$^{-2}$ simulation described in \S~\ref{sec:ideal}. Both panels show results at $t = 60$~fs after the front of the laser pulse has arrived at either the critical density or the perfectly reflecting surface. At this time the standing wave is no longer present and the laser pulse is entirely in reflection, moving towards the bottom of the page.
}\label{fig:collimate}
\end{figure*}

To understand the nature of this difference, we plotted in Fig.~\ref{fig:collimate} the transverse electric fields (i.e. $E_y$) for the $10^{18}$~W~cm$^{-2}$ simulations at $t = 60$~fs after the front of the laser pulse has arrived at the critical surface (or ideal conductor in the idealized simulation). This choice of time highlights the reflected laser field. Also shown in Fig.~\ref{fig:collimate} are electron trajectories from the beginning of the simulation up to $t = 60$~fs. 

Comparing the reflected laser fields in Fig.~\ref{fig:collimate} shows that in the realistic case the laser field is highly modified from interacting with the target whereas in the idealized case the reflected laser pulse is still a simple function of time and space. This essential difference between the reflected laser fields provides a good explanation for why Eq.~\ref{eq:boosted}, which is derived with a simple plane-wave assumption, gives reasonable results for the idealized case but not for the realistic case. 

Comparing the electron trajectories plotted in Fig.~\ref{fig:collimate} for the realistic case and the idealized case we can see how the complex structure of the realistic reflected pulse works to keep escaping electrons closer to the laser axis for significantly longer. Because of the longer interaction time with the laser fields, these electrons ultimately reach higher kinetic energies than in the ideal case. The confinement of these electrons along the laser axis also explains why the realistic simulations feature significant numbers of electrons escaping parallel to the laser axis, whereas the idealized simulations almost exclusively show electrons moving at two preferred angles away from the target, depending on where the electrons originate.

Compared to other studies investigating the reflection of laser pulses from solid targets, \citet{Ruhl_etal1999} attribute the collimation of an electron beam emerging from a solid target at oblique incidence to quasi-static magnetic fields that build up in the pre-plasma. { \citet{Pegoraro_etal1997} also emphasize the importance of quasi-static magnetic fields for electron acceleration in underdense plasmas.} While we see persistent magnetic fields in these PIC simulations after the laser pulse has reflected from the target, careful analysis of escaping electron trajectories indicate strong deflections that are only from the reflected laser pulse.

\begin{figure*}
\includegraphics[angle=0,width=3.2in]{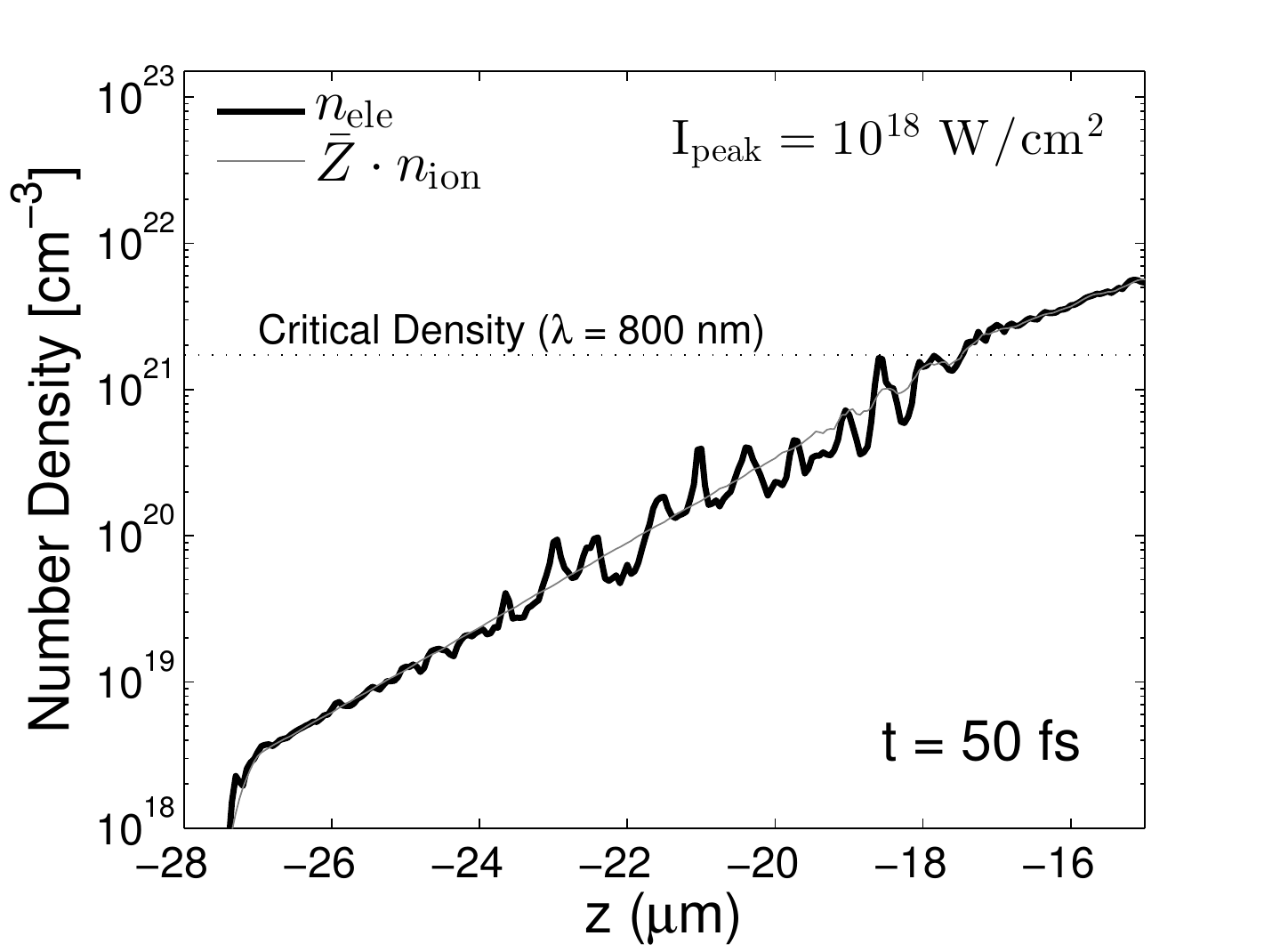}\includegraphics[angle=0,width=3.2in]{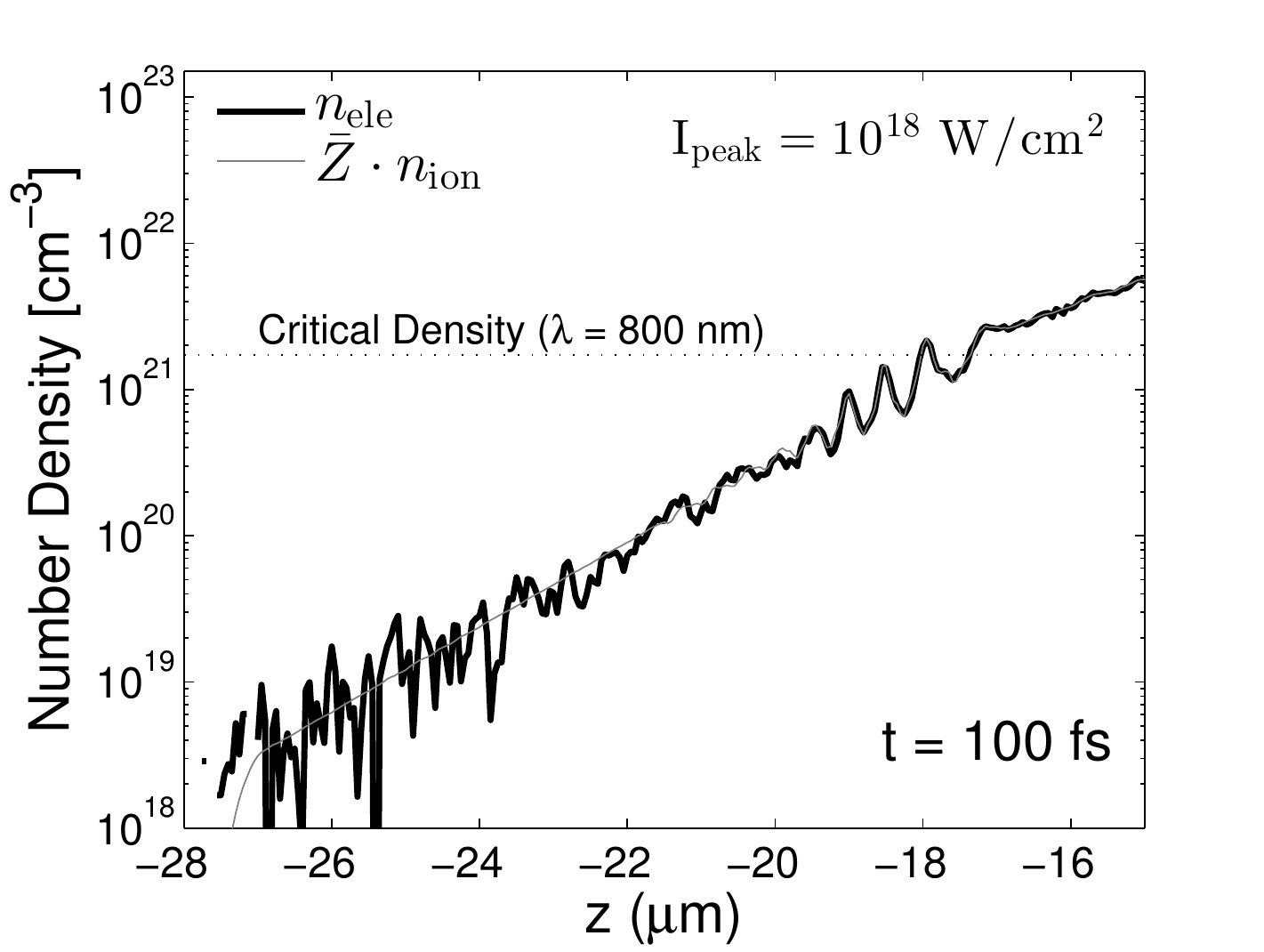}
\vspace{-0.4cm}
\caption{A comparison of the electron number density to the product of the mean ionization state ($\bar{Z}$) and the ion density $n_{\rm ion}$ for the ``realistic'' simulation with $\rm{I}_{\rm peak} = 10^{18}$~W~cm$^{-2}$. Differences in these quantities indicate a charge imbalance and the presence of quasi-static electric fields. The left panel shows results at $t = 50$~fs when the standing wave is present while the right panel shows results at $t = 100$~fs, which is after the standing wave exists in the plasma. The standing wave fields strongly modify the electron density profile near the critical density, creating charge imbalances (left) that modify the ion density profile (right) in a process referred to as ``ponderomotive steepening'' \cite{Estabrook_Kruer1983}. As discussed in the text, the quasi-static electric fields that cause this effect can provide a boost to standing-wave accelerated electrons.} \label{fig:quasistatic}
\end{figure*}

{ We do, however, note the importance of quasi-static \emph{electric} fields. Fig.~\ref{fig:quasistatic} compares the electron density profile along the laser axis to $\bar{Z} \cdot n_{\rm ion}$ in the ``realistic'' $\rm{I}_{\rm peak} = 10^{18}$~W~cm$^{-2}$ simulations at two different times. To the extent that these lines overlap, the plasma is neutral and quasi-static effects are unimportant. The left panel of Fig.~\ref{fig:quasistatic} shows that this is not the case at $t = 50$~fs. Depending on the region, the electron charge density may differ significantly from the charge density of the positive ions. This result is typical during the tens of femtoseconds when the forward and reflected pulses overlap. The electron densities are increased at the nodes of the standing wave, where the transverse electric fields are weak, and suppressed in the anti-nodes of the standing wave, where the transverse electric fields are strong. The right panel of Fig.~\ref{fig:quasistatic} compares $n_{\rm ele}$ and $\bar{Z} n_{\rm ion}$ at $t = 100$~fs, which is a later time when the standing wave fields are not present. Remarkably, by $t = 100$~fs when the plasma is mostly neutral, the ion density profile near critical density has been modified by the charge imbalances created by the standing wave at earlier times. In the literature this process is sometimes called ``ponderomotive steepening'' \cite{Estabrook_Kruer1983}.}

{ Fig.~\ref{fig:quasistatic} shows that the standing wave creates charge imbalances in the pre-plasma producing quasi-static electric fields strong enough to modify the ion densities. Analysis of electron trajectories in the realistic simulations reveal that these quasi-static electric fields can provide an additional boost to electrons that are accelerated away from the target through the standing wave mechanism (Fig.~\ref{fig:sw} or see the next section for how this occurs at $a_0 \sim 0.5$). As described in \S~\ref{sec:sw} these electrons originate from half-way between the nodes and anti-nodes of the electric fields. Interestingly, this is at a position where the electron density is enhanced due to ponderomotive effects \cite{Estabrook_Kruer1983}. The electrons accelerated by the standing wave fields can receive additional energy through repulsion from the overdensity of electrons at the nodes of the electric fields and attraction to the partially-unshielded ions near the anti-node of the electric field. While in some cases the electrostatic forces may oppose the back-directed motion of standing-wave accelerated electrons (which can contribute to the chaotic nature of the trajectories exhibited in Fig.~\ref{fig:nele}), as the standing wave comes to an end this effect can give some electrons increased backward-directed momenta as they are ``injected'' into the reflected laser pulse. Eq.~\ref{eq:boosted} indicates that this could greatly increase the final energies of the launched electrons because it serves to make the denominator of the boost term much smaller. We consider this to be an important factor for why the peak electron energies from the realistic simulations are so much larger than in the ideal simulations where quasi-static electric fields are very small or negligible.}

\subsection{Electron Acceleration at Moderately Relativistic Energies}
\label{sec:mod}

\citet{Kemp_etal2009} treat standing wave acceleration at relativistic intensities ($a_0 \gtrsim 1$), and conclude that $p_{\rm max} = 1.45 \, a_0$ provides a useful rule of thumb when electrons are highly relativistic ($v \sim c$). Appendix~\ref{ap:thresh} provides some approximate analytic insights into standing wave acceleration in this regime and explains how the sequence illustrated in Fig.~\ref{fig:sw} only applies to relativistic electrons. Electrons that are only moderately relativistic ($v \sim 0.5 \, c$, $a_0 \sim 0.5$) cannot be accelerated by this mechanism.

This subsection considers standing wave acceleration for $a_0 \sim 0.5$, which is the relevant $a$-value for the $5 \cdot 10^{17}$~W~cm$^{-2}$ simulations near the laser axis (c.f. Table~\ref{tab:intensities}), and the $10^{18}$~W~cm$^{-2}$ simulations in regions $\sim 1 \mu$m away from the laser axis. Fig.~\ref{fig:er}a indicates that significant numbers of electrons are accelerated even for intensities $a_0 \lesssim 1$, and thus it is important to explain how this occurs.

\begin{figure*}
\includegraphics[angle=0,width=3in]{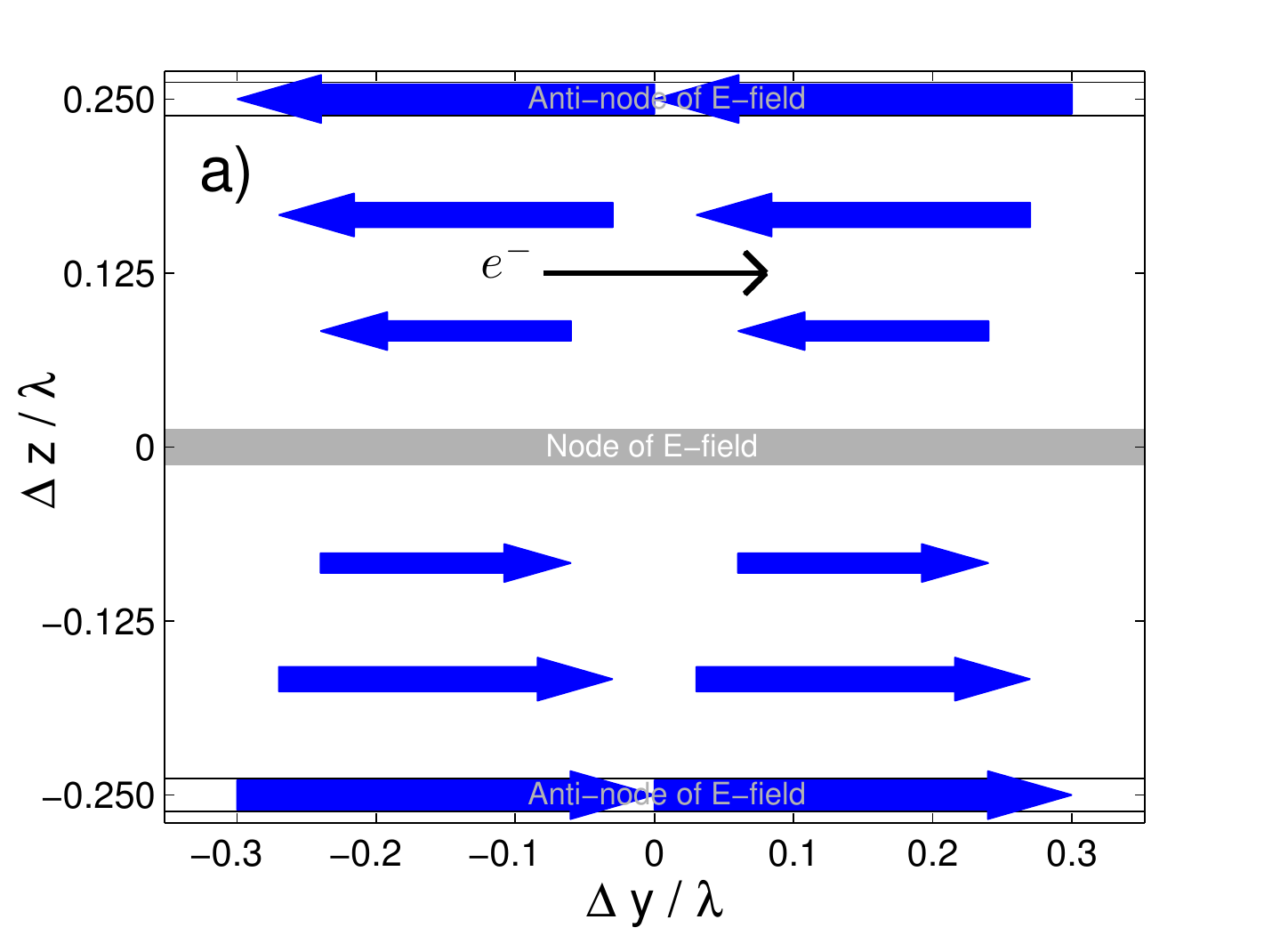}\includegraphics[angle=0,width=3in]{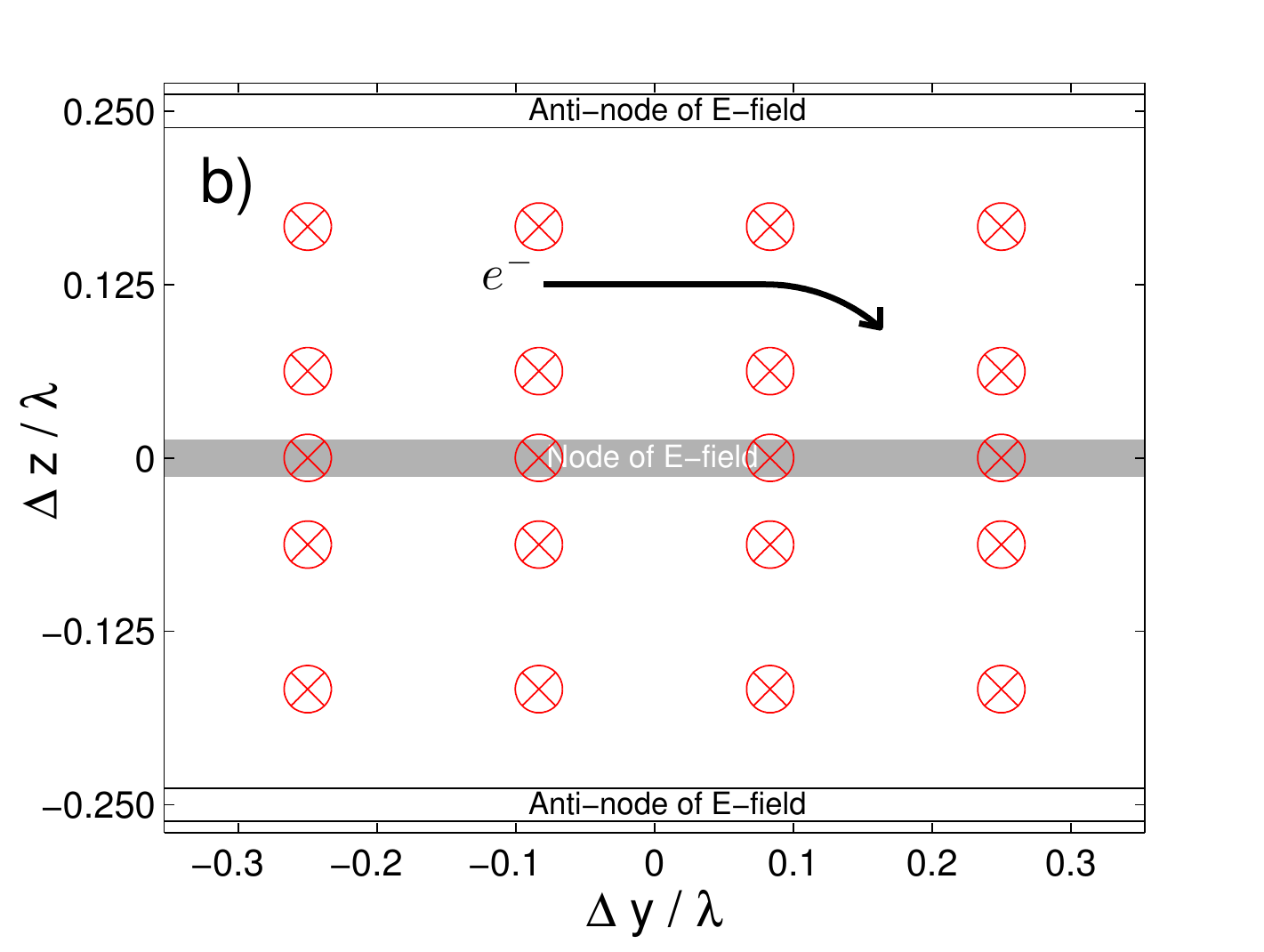}
\includegraphics[angle=0,width=3in]{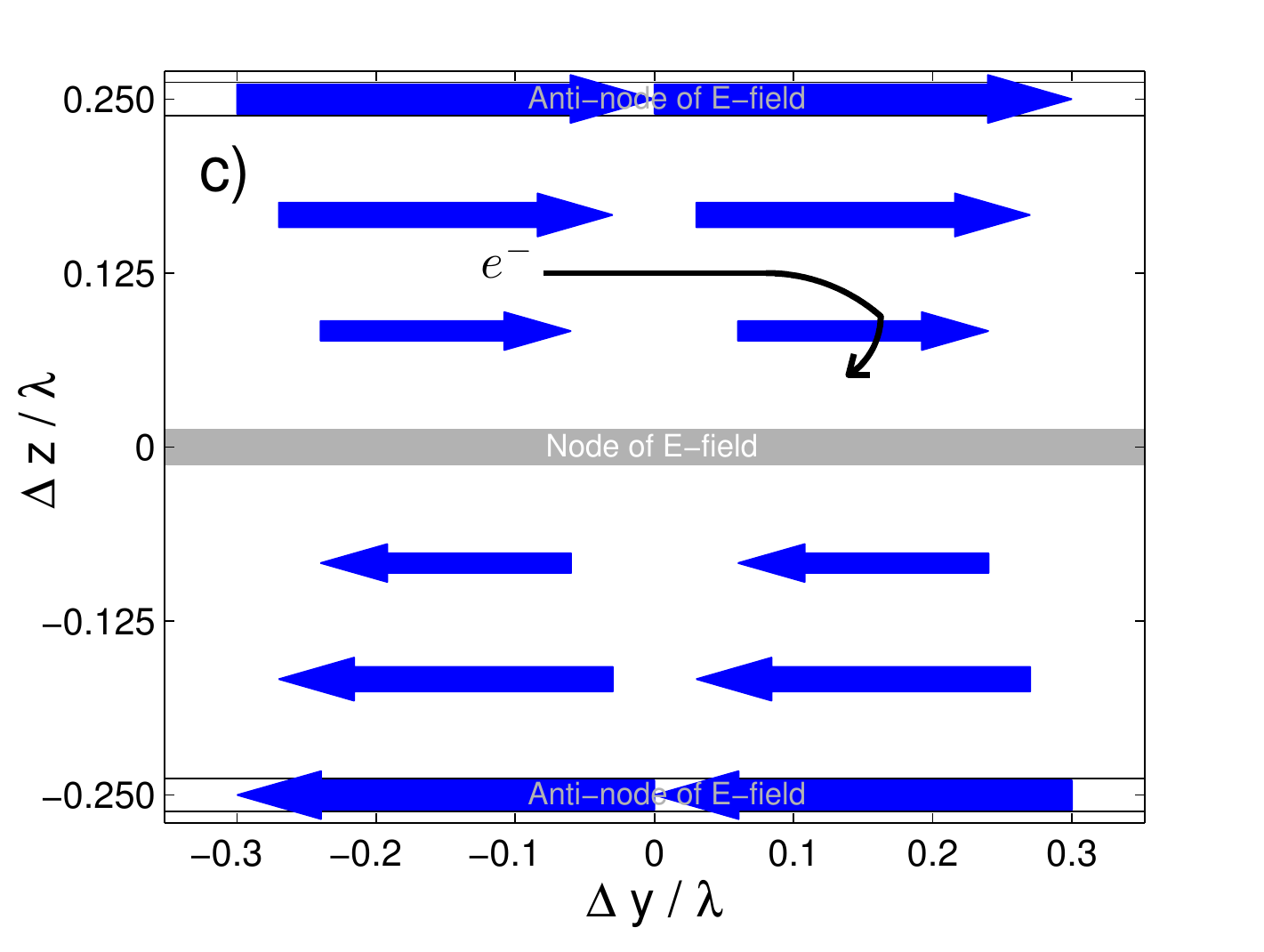}\includegraphics[angle=0,width=3in]{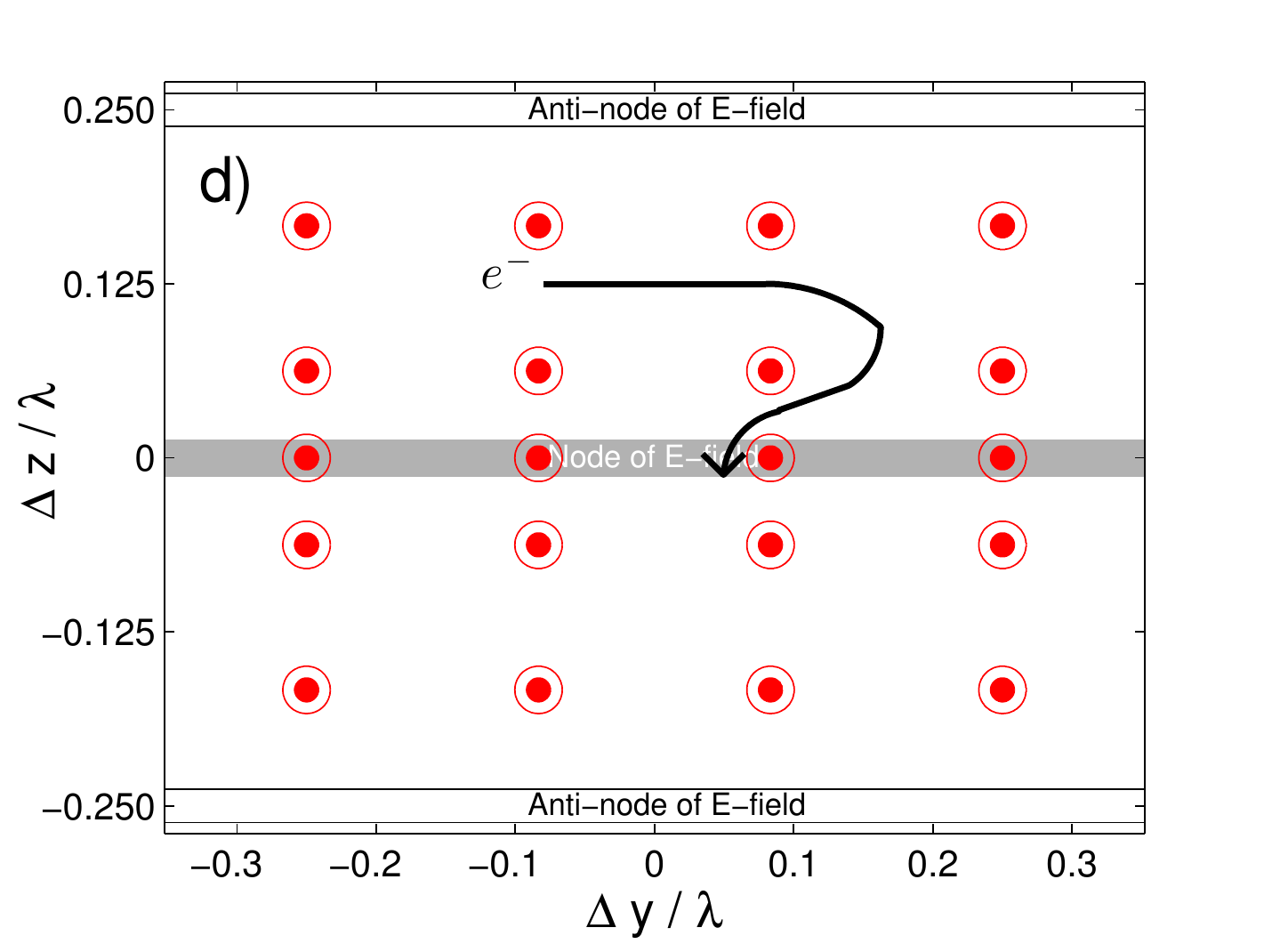}
\caption{An illustration of standing wave acceleration at moderately relativistic intensities ($a_0 \sim 0.5$). Panel a. shows the ``push'' phase, exactly as before. Panel b. illustrates the ``rotate'' phase, which starts the movement away from the target. Panel c. shows how, a quarter cycle later, the standing wave electric fields accelerate the electron to the left. Finally, in Panel d., the electron trajectory can be deflected by the standing wave magnetic fields to become roughly parallel with the laser axis.
}\label{fig:mod}
\end{figure*}

While the electron trajectories in Figs.~\ref{fig:nele} \& \ref{fig:ideal} are highly chaotic inside the standing wave, careful analysis of these trajectories at the moment when the electron begins an overall motion away from the target reveals another pathway for acceleration away from the target besides the sequence illustrated in Fig.~\ref{fig:sw}. This pathway is illustrated in Fig.~\ref{fig:mod}. As before, electrons receive a ``push'' from the laser electric fields in the first step (Panel a), but their velocity is not fast enough to reach the node of the E-field by the end of the ``rotate'' step (Panel b). However, by this point the electron has an appreciable momentum in the $-z$ direction and it will continue to move away from the target during the next step (Panel c) during which the laser electric field accelerates the particle to the left. In the last step (Panel d), the magnetic fields bend the electron trajectory to be nearly parallel with the laser axis, and the electron may continue moving away from the target. Note that this sequence works analogously a half-cycle later for electrons near $\Delta z = 0.125 \lambda$ that are pushed instead to the \emph{left} by the standing wave electric fields followed by an analogous deflection away from the target by the magnetic and electric fields.

Notice that the electron has only moved $\lambda / 8$ away from the target via this mechanism during the course of the laser cycle, whereas the electrons in Fig.~\ref{fig:sw} have moved $\lambda / 4$ away over this same interval. This moderately-relativistic standing wave acceleration is less energetic and less efficient than the relativistic case illustrated in Fig.~\ref{fig:sw}. However, the acceleration depicted in Fig.~\ref{fig:mod} occurs on the timescale of the laser cycle and electrons are ejected in sub-fs bunches, which is desirable for some applications, and these electrons can be launched into the reflected laser pulse and accelerated to substantial energies. To our knowledge this mechanism has not been described in the literature before.

\section{Conclusions and Future Work}
\label{sec:conclusions}

We describe simulations and identify electron acceleration mechanisms relevant to an experiment at the Air Force Research Laboratory in Dayton, OH in which ultra-intense laser pulses ($I_{\rm peak} \approx 10^{18}$ W cm$^{-2}$, $\sim$40 fs FWHM, 3 mJ total energy) are normally incident on a continuous water-jet target. Electrons are ejected from the target and accelerated to relativistic energies through the \emph{reflection} of these pulses from the water jet and when a significant ns-timescale pre-pulse is present. Experimental results are explained in considerable depth in Morrison et al. \cite{Morrison_etal2015}.  Remarkably, the total charge in relativistic electrons is measured to be of order 0.3 nC, which is substantially more charge than comparable laser-wakefield experiments. 

We simulate these laser-matter interactions with 2D(3$v$) PIC simulations using the LSP code and assume an exponential pre-plasma density profile. To investigate mechanisms of electron acceleration, simulations were performed with fixed spot size, temporal duration and pre-plasma scale length but with varying peak intensity. We highlight intensities of $5 \cdot 10^{17}$ W cm$^{-2}$, $10^{18}$ W cm$^{-2}$ and $5 \cdot 10^{18}$ W cm$^{-2}$, which is a regime where the efficiency of electron acceleration increases significantly with increasing intensity.

For comparison, idealized 2D(3$v$) PIC simulations were performed with a very low density pre-plasma and an ideal conductor to reflect the laser light instead of a realistic pre-plasma. Many electrons were accelerated away from the idealized target, as expected, and a combination of simple plane-wave \cite{Yu_etal2000} and standing-wave \cite{Kemp_etal2009} assumptions provided an adequate explanation for the energies of these electrons. Interestingly, these electron energies were generically much \emph{lower} than observed from the realistic targets at the same intensity. The reason for this was associated with the non-ideal nature of the reflected laser pulse { and to quasi-static electric fields created through ponderomotive steepening of the electron density profile \cite{Estabrook_Kruer1983}} in the realistic simulations. Close consideration of electron trajectories within the standing waves created by the forward-going and reflected pulse also revealed a pathway for electron acceleration even when electron velocities are only moderately relativistic ($v \sim 0.5 \, c$).

Having explored electron acceleration mechanisms in some detail, in future work closer connections will be made between simulation and experiment. This will include quantitative comparison of measurements of x-rays emerging from the target chamber and synthetic x-ray spectra predicted by PIC simulations. A number of other diagnostics of the energies and spatial distribution and total charge of the electrons accelerated from the target will also be considered. And, importantly, the assumed pre-plasma densities used in the PIC simulations will become more realistic through use of a novel interferometry system recently installed in the target chamber \cite{Feister_etal2014}. Finally, measured spectra of the reflected light from the target will be compared to analogous measurements of the reflected electromagnetic fields in the simulations. As mentioned previously, we find that, because of the interaction between the laser pulse and the pre-plasma, the reflected laser pulse is substantially modified. Understanding these modifications in simulations and connecting these insights to measurements of the reflected light will be key to understanding and further enhancing the electron energies. This knowledge should ultimately prove useful for a variety of applications.

\section*{Acknowledgements}

CO thanks Douglass Schumacher for insightful conversations and Sheng Jiang for help with MATLAB. CO also thanks Matt Levy for pointing us to \citet{Kemp_etal2009} and Brent Anderson for help with running LSP at AFRL. This research was sponsored by the Quantum and Non-Equilibrium Processes Division of the Air Force Office of Scientific Research, under the management of Dr. Enrique Parra, Program Manager. This project also benefitted from a grant of time at the Spirit supercomputer (AFRL) and storage space at the Ohio Supercomputer Center.

\bibliography{ms}
\bibliographystyle{apsrev}

\appendix

\onecolumngrid

\section{Approximate Intensity Threshold for Standing Wave Acceleration}
\label{ap:thresh}

This section outlines some approximate analytic considerations for the minimum intensity required to achieve standing wave acceleration. It should be emphasized that while the treatment of electron acceleration in this section is approximate, these conclusions have been vetted through careful consideration of electron trajectories in PIC simulations. As one of the main conclusions of this paper, these trajectories indicate that standing wave acceleration can proceed through two different pathways depending on the peak intensity. This was treated in a qualitative way in the body of the paper, and this appendix will provide some analytic arguments for why this occurs.

As will be explained, an intensity threshold can be derived from the requirement that the peak magnetic field strength of the standing wave must be sufficient to deflect electrons away from the target. Another consideration for the minimum intensity arises from a timing requirement. Both of these thresholds are illustrated with vertical dotted lines in Fig.~\ref{fig:er}, and the latter is more stringent. All results in this section come from approximate analytic arguments, beginning with the assumption that Eqs.~\ref{eq:swE} \& \ref{eq:swB} provide a good description of the standing wave electric and magnetic fields. As discussed in \S~\ref{sec:results}, this assumption becomes less accurate at very high intensities ($\gtrsim 5 \cdot 10^{18}$ W cm$^{-2}$). Other assumptions will be made for simplicity.

\subsection{Magnetic Field Strength Required for Deflection}

First, we consider the requirement on the magnetic field strength. In what follows the electron is assumed to start from rest and, as in other parts of this paper, motion towards the target is in the $z$ direction whereas motion parallel to the target is in the $\pm y$ direction. The electron starts from rest at a position half-way between the node and the anti-node in the E field, as depicted in Fig.~\ref{fig:sw}.

During the ``push'' phase, the $y$-momentum of the electron increases significantly. If we ignore the magnetic fields during this phase, the changing $y$-momentum is given by (Eq.~\ref{eq:swE})
\begin{equation}
\frac{d p_y}{ dt} \approx q E_{y}(t) \approx \frac{2}{\sqrt{2}} \, q  E_{y0} \sin(\omega t). \label{eq:push}
\end{equation}
where in the last step the $z$-dependent factor in Eq.~\ref{eq:swE} was approximated as $1/\sqrt{2}$ because of the electron's location half-way between the node and anti-node of the standing wave electric field. Eq.~\ref{eq:push} can be integrated over a half cycle to estimate the momentum gained during the ``push'' phase,
\begin{equation}
\Delta p_y = \int_0^{\pi/\omega} \left( \frac{dp_y}{dt}\right) \, dt = \sqrt{2} \, q E_{y0} \int_0^{\pi/\omega} \sin (\omega t) dt = \frac{\sqrt{2}}{\pi} \, q \, E_{y0} \tau_p \label{eq:deltapy}
\end{equation}
where the optical period in the lab frame is $\tau_p = 2 \pi / \omega$. This expression will be used later to determine how far an electron moves during a half-cycle.

During the ``rotate'' phase, the electron will acquire a non-zero momentum away from the target ($-z$). This momentum can be estimated in an analogous way by neglecting the electric fields and just considering the standing wave magnetic fields. In this case our equations of motion (in 2D) are
\begin{equation}
\frac{dp_z}{dt} \approx -\gamma q v_y(t) B_x(t) = -\frac{q}{m} \, B_x (t) \, p_y (t) \label{eq:dpz}
\end{equation}
\begin{equation}
\frac{dp_y}{dt} \approx \gamma q v_z (t) B_x(t) = \frac{q}{m} \, B_x (t) \, p_z (t).
\end{equation}
Since electric fields are being neglected in this approximate treatment of the ``rotate'' phase, the electron cannot actually gain energy and $p_y$ and $p_z$ are subject to the constraint:
\begin{equation}
p_0 = \sqrt{p_y^2 + p_z^2} = {\rm constant}. \label{eq:p0}
\end{equation}
Since $B_x(t)$ is specified by Eq.~\ref{eq:swB}, there are only two unknowns, $p_y$ and $p_z$, and only two of the previous three equations are needed. Solving for $p_y$ in Eq.~\ref{eq:p0}, and substituting in Eq.\ref{eq:dpz} yields
\begin{equation}
\frac{dp_z}{dt} = - \frac{q}{m} \, B_x(t) \, \sqrt{p_0^2-p_z^2}.
\end{equation}
An estimate for the minimum standing wave magnetic field strength can be derived from this expression by integration,
\begin{equation}
\int \frac{dp_z}{\sqrt{p_0^2-p_z^2}} = -\frac{q}{m} \int B_x(t) \, dt. \label{eq:int}
\end{equation}
Now the limits of integration must be specified. Assuming that the magnetic field strength is sufficient to rotate the particle from an initial state moving parallel to the target ($p_y = p_0$, $p_z = 0$), to a state where the particle is moving directly away from the target ($p_y = 0$, $p_z = p_0$), then the limits of integration on the $dp_z$ integral must be zero and $p_0$,
\begin{equation}
\int_0^{p_0} \frac{dp_z}{\sqrt{p_0^2-p_z^2}} = \tan^{-1} (\infty) = \frac{\pi}{2}.
\end{equation}
For a 90-degree rotation to occur the integral over the $-q B_x (t)$ term in Eq.~\ref{eq:int} must be greater than or equal to $\pi / 2$. The $-q B_x(t)$ term will be integrated over a half-cycle,
\begin{equation}
-\frac{q}{m} \int_{\pi / 2 \omega}^{3 \pi / 2 \omega} B_x(t) dt = - \frac{q}{m}  \sqrt{2} \, B_{x0} \int_{\pi / 2 \omega}^{3 \pi / 2 \omega} \cos(\omega t) dt = \frac{\sqrt{2}}{\pi} \frac{q}{m} B_{x0} \tau_p
\end{equation}
The requirement on the magnetic field is therefore
\begin{equation}
B_{x0} \gtrsim \frac{\pi^2}{2 \sqrt{2} \left( \frac{q}{m} \right) \tau_p  } \approx 7,400 \, {\rm T}
\end{equation}
where the numerical value comes from assuming $\lambda = 800$ nm light ($\rightarrow \tau_p = 2.67$ fs). Using this numerical value for $B_{x0}$ implies an electric field of $E_{y0} = c B_{x0} = 2.2 $\, TV / m  and a corresponding intensity,
\begin{equation}
I_{\rm thresh} = \frac{c \epsilon_0 E_{y0}^2}{2} = 6.6 \cdot 10^{17} \, {\rm W} \, {\rm cm}^{-2}. \label{eq:thresh}
\end{equation}
This ``threshold'' should be understood as the intensity needed for electrons to acquire a significant $p_z$ momentum towards or away from the target. As will soon be discussed, somewhat higher intensities than this are required to produce the electron trajectories described in Fig.~\ref{fig:sw}.

\subsection{Timing Considerations}

In order for the standing wave acceleration mechanism to work as described in Fig.~\ref{fig:sw}, it is very important that electrons reach the node of the electric field by the end of the ``rotate'' phase (Panel b). Thus, at later times when the standing wave electric fields are strong (Panel c) the electron is positioned where it can continue ``drifting'' away from the target. Electrons must travel from their initial position to the node of the electric field ($|\Delta z| = \lambda / 8$). This motion is approximately circular with radius $\lambda / 8$ so the quarter-circumference distance traveled is $(2 \pi / 4 ) \cdot (\lambda / 8)$.  Most of this motion occurs during the one-\emph{quarter} of the laser period when the electron has an appreciable velocity transverse to the target \emph{and} the standing wave magnetic fields are dominant. The effective speed, $v_{\rm eff}$, of the electron during the ``rotate'' phase must therefore be
\begin{equation}
v_{\rm eff} = \frac{(2 \pi / 4) (\lambda / 8) }{\tau_p / 4} = \frac{\pi}{4} \frac{\lambda}{\tau_p} = \frac{\pi}{4} c \, \approx \, 0.79 \, c.
\end{equation}
At the threshold intensity, the momentum at the \emph{end} of the ``push'' phase, according to Eq.~\ref{eq:deltapy}, corresponds to a velocity of $0.84 c$. The ``effective'' speed would be somewhat less than this. The ``push'' and ``rotate'' phases overlap with each other because of the $\cos (\omega t)$ and $\sin( \omega t)$ terms in Eqs.~\ref{eq:swE} \& \ref{eq:swB}. At the time when the standing wave magnetic field becomes dominant the instantaneous speed of the electron (calculated using different limits in Eq.~\ref{eq:deltapy}) would be closer to $0.63 c$. And the true value would be slightly \emph{less} than this because, although the electron is moving closer to the node of the standing wave electric field, the spatial ($z$-dimension) dependence of Eq.~\ref{eq:swE} was ignored in Eqs.~\ref{eq:push} \& \ref{eq:deltapy}. 

In the end one finds that significantly greater intensities ($\gtrsim 10^{18}$ W cm$^{-2}$) than the threshold intensity ($6.6 \cdot 10^{17}$ W cm$^{-2}$) are required to ensure $v_{\rm eff} \sim c$ so that the electron reaches the node of the standing wave electric field in adequate time. The threshold intensity is still a useful indicator for when standing wave magnetic fields may be strong enough to provide some deflection away from the target. As an example of this, the so-called ``moderately relativistic'' case is treated in \S~\ref{sec:mod} and illustrated in Fig.~\ref{fig:mod} as part of an explanation for why PIC simulations with $5 \cdot 10^{17}$ W cm$^{-2}$ peak intensities still indicate significant numbers of electrons leaving the target. As \S~\ref{sec:mod} discusses, this pathway to electron acceleration is qualitatively different than at significantly higher intensities.

\end{document}